\documentstyle[aps,preprint]{revtex}
\begin{document}
\draft{}
\title{ Search for the scalar $a_0$ and $f_0$ mesons in the reactions
$e^+e^-\to\gamma\pi^0\pi^0(\eta)$.}

\author{N.N. Achasov
\thanks{ E-mail: achasov@math.nsc.ru} and  V.V. Gubin
\thanks{ E-mail: gubin@math.nsc.ru }}
\address{Laboratory of Theoretical Physics\\
S.L. Sobolev Institute for Mathematics\\
630090 Novosibisk 90,\  Russia}
\date{\today}
\maketitle
\begin{abstract}

It is shown that the  reactions $e^+e^-\to\gamma\pi^0\pi^0(\eta)$ give a good
chance for observing the scalar $a_0$ and $f_0$ mesons.
In the photon energy region less then $100$ MeV the vector meson
contributions $e^+e^-\to V^0\to\pi^0 V'^0\to\gamma\pi^0
\pi^0(\eta)$ are negligible in comparison with the scalar meson
 $e^+e^-\to\phi\to\gamma f_0(a_0)\to\gamma\pi^0\pi^0(\eta)$
for  $BR(\phi\to\gamma f_0(a_0)\to\gamma\pi^0\pi^0(\eta))$
greater than $5\cdot10^{-6}(10^{-5})$.
Using the two-channel treatment of the $\pi\pi$ scattering the predictions
for $BR(\phi\to\gamma (f_0+\sigma)\to\gamma\pi\pi)$ are derived. The
 four quark model, the model of $K\bar K$ molecule
and the $s\bar s$ model of scalar $f_0$ and $a_0$ mesons are discussed.

\end{abstract}


\section{ Introduction.}

  The central problem of light hadron spectroscopy has been the problem
of the scalar $f_0(980)$ and $a_0(980)$ mesons. It is well known that these
states possess peculiar properties from the naive quark ($q\bar q$)
model point of view, see, for example, the reviews
\cite{{montanet},{achasov-84},{close-88},{landsberg},{achasov-91}}.
 To clarify the nature of these mesons a number
of models has been suggested.
It was shown that all their challenging properties
could be understood \cite{achasov-84,achasov-91,achasov-1991}
in the framework of the four-quark  ($q^2\bar q^2$) MIT-bag model
 \cite {jaffe-77}, with symbolic quark structure 
 $f_0(980)=s\bar s(u\bar u+d\bar d)/
\sqrt{2}$ and $a_0(980)=s\bar s(u\bar u-d\bar d)/\sqrt{2}$. Along with the
 $q^2\bar q^2$ nature of  $a_0(980)$ and $f_0(980)$ mesons the possibility of
 their being the  $K\bar K$ molecule is discussed \cite {weinstein-90}.

During the last few years it was established
\cite{achasov-89,isgur-93,molecule,achasov-95} that the radiative decays
 of the $\phi$ meson $\phi\rightarrow\gamma f_0\rightarrow\gamma\pi\pi$ and
$\phi\rightarrow\gamma a_0\rightarrow\gamma\eta\pi$
could be a good guideline in distinguishing  the $f_0$ and $a_0$ meson
models. The branching ratios are considerably different in the cases of
naive quark, four-quark or molecular models. As has been shown
\cite{achasov-89}, in the four quark model the branching ratio is
\begin{equation}
BR(\phi\to\gamma f_0(q^2\bar q^2)\to\gamma\pi\pi)\simeq
BR(\phi\to\gamma a_0(q^2\bar q^2)\to\gamma\pi\eta)\sim10^{-4},
\end{equation}
and in the $K\bar K$ molecule model it is \cite{{isgur-93},{molecule}}
\begin{equation}
BR(\phi\to\gamma f_0(K\bar K)\to\gamma\pi\pi)\simeq
BR(\phi\to\gamma a_0(K\bar K)\to\gamma\pi\eta)\sim10^{-5}.
\end{equation}

 Currently also an interest in an old interpretation of the $f_0$
meson being an $s\bar s$ state, see, \cite{tornqvist1,tornqvist2} is
rekindled, despite the fact that the almost ideal mass degeneracy of the
 $f_0$ and  $a_0$ mesons is difficult to understand in this case.
By adding the quark-gluon transitions  $q\bar q\leftrightarrow gg$ one
does not manage  this problem. The experiment points rather to the fact
that the $f_0$ meson is weakly coupled with gluons. Really, according QCD
 \cite{qcd} we have
\begin{equation}
BR(J/\psi\to\gamma+gg \mbox{ in\ $0^+$\ or\ $0^-$\ states })=1,5\cdot10^{-2},
\end{equation}
but from experiment \cite{eigen}
\begin{equation}
BR(J/\psi\to\gamma gg\to\gamma f_0\to\gamma\pi\pi)<1,4\cdot10^{-5}
\end{equation}
and \cite{particle-94}
\begin{equation}
BR(J/\psi\to\gamma gg\to\gamma\eta'(958))=4,3\cdot10^{-3}.
\end{equation}

Hence, if one considers that the $\eta'(958)$ coupling with
 gluons is essential,
then the $f_0(980)$ coupling with the gluons should be considered as rather
weak.

In spite of this fact, the  $s\bar s$ scenario is discussed in the current
literature as one possible model of the $f_0$ meson structure. The
"accidental" mass degeneracy of $a_0$ and $f_0$ mesons in this case  is
assigned to the final state interaction.

It is easy to note that in the case of an $s\bar s$ structure
 of the $f_0$ meson
$BR(\phi\to\gamma f_0\to\gamma\pi\pi)$ and
$BR(\phi\to\gamma a_0\to\gamma\pi\eta)$ are different by factor of ten,
wich should be visible experimentally.
Really, the  $a_0$ meson as an isovector state has a symbolic structure
$a_0=(u\bar u-d\bar d)/\sqrt{2}$ in the two-quark model. In this case the
 decay $\phi\to\gamma a_0\to\gamma\pi\eta$ is suppressed by the OZI rule
and the rough estimation for the real part of the decay amplitude gives
\cite{achasov-95} $BR(\phi\to\gamma a_0(q\bar q)\to\gamma\pi\eta)
\sim3\cdot10^{-7}$.

By virtue of the real two-particle $K^+K^-$ intermediate
state the OZI rule breaking imaginary part of the decay amplitude  is
relatively high and gives \cite{achasov-89}
\begin{equation}
BR(\phi\to\gamma a_0(q\bar q)\to\gamma\pi\eta)\simeq5\cdot10^{-6}.
\end{equation}

 In the case when $f_0=s\bar s$ the suppression by the OZI rule is absent
and the evaluation gives
\cite{achasov-89}
\begin{equation}
BR(\phi\to\gamma f_0(s\bar s)\to\gamma\pi\pi)\simeq5\cdot10^{-5}.
\end{equation}
Let us note that in the case of the $\phi\to\gamma\eta'$ decay allowed by the
OZI rule one expects  $BR(\phi\to\gamma \eta')\simeq10^{-4}$.

 So, the decays $\phi\rightarrow\gamma f_0\rightarrow\gamma\pi\pi$ and
$\phi\rightarrow\gamma a_0\rightarrow\gamma\eta\pi$ are of special interest
for both theorists and the experimentalists.

At the present time the investigation of the  $\phi\rightarrow\gamma
f_0\rightarrow\gamma\pi^+\pi^-$  decay has started with the detector CMD-2
\cite{novo} at  the $e^+e^-$-collider VEPP-2M in Novosibirsk.
Besides that, in Novosibirsk at the same collider the detector SND has started
 \cite{snd} and  has been working  with
$e^+e^-\to\gamma f_0\rightarrow\gamma\pi^0\pi^0$ and  $e^+e^-\to\gamma a_0
\rightarrow\gamma\eta\pi^0$ decays. The modernization
of the VEPP-2M complex is planed to increase the luminosity
by one order of magnitude \cite{particle-94}. And,
finally, in the near future in Frascati the start of the operation of
the  $\phi$-factory DA$\Phi$NE is expected  \cite{particle-94}, which
should make possible extensive studies of the scalar $f_0(980)$ and $a_0(980)$
  mesons.

We have shown recently \cite{interference} that  the search for the $f_0$
meson in the reaction $e^+e^-\to\phi\to\gamma f_0\to\gamma\pi^+\pi^-$
is not an easy task because of the large initial state radiation background.
 In this paper we study the reactions $e^+e^-\to\phi\rightarrow\gamma
 (f_0+\sigma)\rightarrow\gamma\pi^0\pi^0$ and
 $e^+e^-\to\phi\rightarrow\gamma a_0\rightarrow\gamma\pi^0\eta$.

In the second part of the paper, imposing the appropriate photon energy
cuts $\omega<100$ MeV, we show that the background reactions
$e^+e^-\to\rho(\omega)\to\pi^0\omega(\rho)\to\gamma\pi^0\pi^0$,
$e^+e^-\to\rho(\omega)\to\pi^0\omega(\rho)\to\gamma\pi^0\eta$ and
$e^+e^-\to\phi\to\pi^0\rho\to\gamma\pi^0\pi^0(\eta)$ are negligible up to
$BR(\phi\to\gamma f_0(a_0)\to\gamma\pi^0\pi^0(\eta),\omega<100\ MeV)
 \sim5\cdot10^{-6}(10^{-5})$. We assume that it will be experimentally 
 possible to isolate photons from $\phi\to\gamma f_0$ and $\phi\to\gamma a_0$
 with energies $\omega<100$ MeV, despite the background of the
 other photons from decays of $\pi^0$ and $\eta$. This cut plays a
 strong role in  suppression of background, see Sec.II below.

In the third part, basing on a two-channel  analysis of the $\pi\pi$
scattering the predictions on $BR(\phi\to\gamma (f_0+\sigma)\to\gamma
\pi\pi)$ are made. We discuss the four-quark model of the  $f_0$ and $a_0$
mesons, the model of the $K\bar K$ molecules and the model of the $f_0$
meson being the $s\bar s$ state as well.

The fourth part is devoted to discussions of the obtained results.

\section{Background to the  \lowercase{$e^+e^-\to\gamma f_0(a_0)
\to\gamma \pi^0\pi^0(\eta)$} reaction in the vector meson dominance model.}

Let us consider the background to the $e^+e^-\to\gamma f_0\to\gamma
\pi^0\pi^0$ reaction. In the vector meson dominance model the diagrams are
pictured in Fig.1. Let us estimate the cross section of the
$e^+e^-\to\rho\to\pi^0\omega\to\pi^0\pi^0\gamma$ process, shown in Fig.1(a).
 First of all, notice that
\begin{eqnarray}
\label{fullrho}
&&\sigma(e^+e^-\to\rho\to\pi^0\omega\to\pi^0\pi^0\gamma,\ m_{\phi})=
\\ \nonumber
&&=\sigma(e^+e^-\to\rho\to\pi^0\omega,\ m_{\phi})\frac{2}{\pi}
\int_{m_{min}}^{m_{max}}\frac{m^2
\Gamma_{\omega\pi\gamma}(m)\Lambda^{3/2}_{\omega}f_{\omega}(m)dm}
{|m_{\omega}^2-m^2-i\Gamma_{\omega}m|^2},
\end{eqnarray}
where
\begin{equation}
\Gamma_{\omega\pi\gamma}(m)=\Gamma_{\omega\pi\gamma}(m_{\omega})
\left(\frac{m_{\omega}}{m}\right)^3\frac{(m^2-m_{\pi}^2)^3}{(m_{\omega}^2-
m_{\pi}^2)^3},\  \Lambda_{\omega}=\frac{(m_{\phi}^2-(m_{\pi}-m)^2)
(m_{\phi}^2-(m_{\pi}+m)^2)}{(m_{\phi}^2-(m_{\pi}-m_{\omega})^2)
(m_{\phi}^2-(m_{\pi}+m_{\omega})^2)},
\end{equation}
and $m$ is the invariant mass of the $\pi\gamma$ system. The limits of the
integration are taken with regard to the photon energy  cuts
$m_{min}\simeq\sqrt{m_{\pi}^2+2m_{\pi}^2\omega_{min}/m_{\phi}}=m_{\pi}$ and
$m_{max}\simeq\sqrt{m_{\pi}^2+2m_{\phi}\omega_{max}}=0.471\ GeV$.
The function $f_{\omega}(m)$ takes into account the interference of the
identical pions, see Appendix.

The cross section of the $e^+e^-\to\rho\to\pi^0\omega$ process
\begin{equation}
\sigma(e^+e^-\to\rho\to\pi^0\omega,s)=12\pi a
\frac{\Gamma(\rho\to e^+e^-,s)\Gamma(\rho\to\pi^0
\omega,s)}{|D_{\rho}(s)|^2},
\end{equation}
where $\Gamma(\rho\to e^+e^-,s)$ is defined in Eq.(\ref{elwidth})  and
\begin{equation}
\Gamma(\rho\to\pi^0\omega,s)=\frac{g^2_{\rho\pi\omega}}{96\pi s\sqrt{s}}
[(s-(m_{\omega}-m_{\pi})^2)(s-(m_{\omega}+m_{\pi})^2)]^{3/2}
\end{equation}
The coupling constant $g_{\rho\pi\omega}$ is taken from the width of the
decay $\omega\to\pi\gamma$ data that leads to  $g^2_{\rho\pi\omega}/96\pi=
0.452$ $GeV^{-2}$.
In the propagator  $D_{\rho}$ we are taking into account the energy
dependence of the $\rho$ meson width
\begin{equation}
D_{\rho}(s)=m_{\rho}^2-s-is\frac{g^2_{\rho\pi\pi}}{48\pi}(1-
\frac{4m_{\pi}^2}{s})^{3/2}.
\end{equation}

The constant $a$ is taken from the condition
$\sigma(e^+e^-\to\pi^0\omega,\sqrt{s}=0.97)=6.1$ nb, see \cite{dol},
that leads to  $a=2.02$.

From the above, the cross section $\sigma(e^+e^-\to\rho\to\pi^0\omega,
m_{\phi})=6.46$ nb. The cross section of the
 $e^+e^-\to\rho\to\pi^0\omega\to\pi^0\pi^0\gamma$ process is
\begin{equation}
\label{rho}
\sigma(e^+e^-\to\rho\to\pi^0\omega\to\pi^0\pi^0\gamma, m_{\phi})=
6.1\cdot10^{-4}\ nb.
\end{equation}
in the region of interest $0<\omega<100$ MeV. Analogously let us consider
the $e^+e^-\to\omega\to\pi^0\rho \to\gamma\pi^0\pi^0$ process, see Fig.1(b).
 First, we have
\begin{eqnarray}
\label{fullomega}
&&\sigma(e^+e^-\to\omega\to\pi^0\rho\to\pi^0\pi^0\gamma,\ m_{\phi})=
\\ \nonumber
&&=\sigma(e^+e^-\to\omega\to\pi^0\rho,\ m_{\phi})\frac{2}{\pi}
\int_{m_{min}}^{m_{max}}\frac{m^2
\Gamma_{\rho\pi^0\gamma}(m)\Lambda^{3/2}_{\rho}f_{\rho}(m)dm}
{|D_{\rho}(m)|^2},
\end{eqnarray}
where
\begin{equation}
\Gamma_{\rho\pi^0\gamma}(m)=\Gamma_{\rho\pi^0\gamma}(m_{\rho})
\left(\frac{m_{\rho}}{m}\right)^3\frac{(m^2-m_{\pi}^2)^3}{(m_{\rho}^2-
m_{\pi}^2)^3},\  \Lambda_{\rho}=\frac{(m_{\phi}^2-(m_{\pi}-m)^2)
(m_{\phi}^2-(m_{\pi}+m)^2)}{(m_{\phi}^2-(m_{\pi}-m_{\rho})^2)
(m_{\phi}^2-(m_{\pi}+m_{\rho})^2)}.
\end{equation}
The cross section of the $e^+e^-\to\omega\to\pi^0\rho^0$ process is
\begin{equation}
\sigma(e^+e^-\to\omega\to\pi^0\rho,s)=12\pi b
\frac{\Gamma(\omega\to e^+e^-,s)\Gamma(\omega\to\pi^0\rho,s)}
{|D_{\omega}(s)|^2}.
\end{equation}

The constant $b$ is found from the condition
$\sigma(e^+e^-\to\omega\to\pi^0\pi^+\pi^-; \sqrt{s}=1.09)\simeq
\sigma(e^+e^-\to\omega\to\pi^0\rho; \sqrt{s}=1.09)\simeq 2.4$ nb, see
\cite{dol}, that leads to $b=4.7$. In view of it the cross section
$\sigma(e^+e^-\to\omega\to\pi^0\rho; m_{\phi})\simeq2.2$ nb. For the cross
section of the $e^+e^-\to\omega\to\pi^0\rho\to\gamma\pi^0\pi^0$ process in
the photon energy region $0<\omega<100\ MeV$, see (\ref{fullomega}),
 one has
\begin{equation}
\label{omega}
\sigma(e^+e^-\to\omega\to\pi^0\rho\to\gamma\pi^0\pi^0, m_{\phi})
= 3.2\cdot10^{-5}\ nb.
\end{equation}

The cross section of the $e^+e^-\to\phi\to\pi^0\rho^0\to\gamma\pi^0\pi^0$
process, see  Fig.1(c), in the $\phi$ meson region at $0<\omega<100$ MeV is
\begin{eqnarray}
\label{fullphi}
&&\sigma(e^+e^-\to\phi\to\pi^0\rho\to\gamma\pi^0\pi^0, m_{\phi})=
\\ \nonumber
&&=\sigma(e^+e^-\to\phi\to\pi^0\rho, m_{\phi})\frac{2}{\pi}
\int_{m_{min}}^{m_{max}}\frac{m^2
\Gamma_{\rho\pi^0\gamma}(m)\Lambda^{3/2}_{\rho}f_{\rho}(m)dm}
{|D_{\rho}(m)|^2}<2.6\cdot10^{-3}\ nb,
\end{eqnarray}
where $BR(\phi\to\pi^0\rho^0\to\gamma\pi^0\pi^0,
\omega<100\ MeV)<6.4\cdot10^{-7}$ \footnote{ Taking into account the positive
interference one has $BR(\phi\to\pi^0\rho^0)<(1/3)BR(\phi\to\pi\rho)$.}.

In principle, the interference between the amplitudes $e^+e^-\to V\to\gamma
V'\to\gamma\pi^0\pi^0$ may be essential, but, as it is seen from
Eq.(\ref{rho}), Eq.(\ref{omega}) and Eq.(\ref{fullphi}), in studies of the
 background to the $e^+e^-\to\gamma f_0\to\gamma\pi^0\pi^0$ reaction one
can neglect the contributions of the $e^+e^-\to\rho(\omega)\to\pi^0\omega
(\rho)\to\pi^0\pi^0\gamma$ processes and the question of their interference
no longer arises.

Hence, taking into account that $\sigma(e^+e^-\to\phi\to all)=
4.4\cdot10^3\ nb$, the background processes
 $e^+e^-\to V^0\to\pi^0 V'^0\to\gamma\pi^0
\pi^0$ are negligible  compared to the one under study  
 $e^+e^-\to\gamma f_0\to\gamma\pi^0\pi^0$ 
 up to $BR(\phi\to\gamma f_0\to\gamma\pi^0\pi^0)=5\cdot10^{-6}$
\footnote{ Let us emphasize that we essentially use the photon energy cuts
in our analyses. For comparison we notice that after integration over
all spectrum $BR(\phi\to\pi^0\rho^0\to\gamma\pi^0\pi^0)\simeq10^{-5}$
\cite{achasov-89,bramon,franzini,tuan}.}.

 As for the interference between
the primary process $e^+e^-\to\gamma f_0\to\gamma\pi^0\pi^0$ and the 
background process $e^+e^-\to\phi\to\pi^0\rho\to\gamma\pi^0\pi^0$, 
one can neglect it also due to the fact that  66\% of background 
contribution to the branching ratio under consideration is determined by the 
region $70<\omega<100$ MeV while the contribution of the primary 
process in this region is less than 20\% in the worst case when
the $f_0$ meson is relatively wide, see Sec.IV.

 Let us consider the background to the  $e^+e^-\to\gamma a_0\to\gamma\pi^0
\eta$ reaction. 
 It is easy to show that in this case, as in the case of the reaction
$e^+e^-\to\gamma f_0\to\gamma\pi^0\pi^0$,
the process $e^+e^-\to\phi\to\pi^0\rho\to\gamma\pi^0\eta$ is 
dominant. The cross sections of the processes 
 $e^+e^-\to\rho(\omega)\to\eta\rho(\omega)\to\gamma\pi^0
\eta$, see Fig.2, are by two orders of magnitude less then the cross 
sections of the processes
 $e^+e^-\to\rho(\omega)\to\pi^0\omega(\rho)\to\gamma\pi^0
\eta$. The cross sections of the processes 
$e^+e^-\to\rho(\omega)\to\pi^0\omega(\rho)\to\gamma\pi^0\eta$, in turn,
are  by two orders of magnitude less then the cross section of 
the process $e^+e^-\to\phi\to\pi^0\rho\to\gamma\pi^0\eta$. 
In comparison with  the  process 
$e^+e^-\to\phi\to\pi^0\rho\to\gamma\pi^0\eta$ the cross section of the 
process  $e^+e^-\to\phi\to\eta\phi\to\gamma\pi^0\eta$ is suppressed by
two orders of magnitude also.

The cross section of the process 
$e^+e^-\to\phi\to\pi^0\rho\to\gamma\pi^0\eta$
\begin{eqnarray}
&&\sigma(e^+e^-\to\phi\to\pi^0\rho\to\pi^0\eta\gamma, m_{\phi})=
\\ \nonumber
&&=\sigma(e^+e^-\to\phi\to\pi^0\rho, m_{\phi})\frac{2}{\pi}
\int_{m_{min}}^{m_{max}}\frac{m^2\Gamma_{\rho\eta\gamma}(m)
\Lambda^{3/2}_{\rho}dm}{|D_{\rho}|^2}.
\end{eqnarray}
Let us take into account that 
\begin{equation}
\frac{2}{\pi}\int_{m_{min}}^{m_{max}}\frac{m^2
\Gamma_{\rho\eta\gamma}(m)\Lambda^{3/2}_{\rho}dm}
{|D_{\rho}|^2}=3.5\cdot10^{-5},
\end{equation}
where
\begin{equation}
\Gamma_{\rho\eta\gamma}(m)=\Gamma_{\rho\eta\gamma}(m_{\rho})
\left(\frac{m_{\rho}}{m}\right)^3\frac{(m^2-m_{\eta}^2)^3}{(m_{\rho}^2-
m_{\eta}^2)^3},
\end{equation}
and $m$ is the invariant mass of the $\eta\gamma$ system.
The integration limits are taken with regard to the photon energy cuts
$m_{min}\simeq\sqrt{m_{\eta}^2+2m_{\pi}^2\omega_{min}/m_{\phi}}
=m_{\eta}$ and
$m_{max}\simeq\sqrt{m_{\eta}^2+2m_{\phi}\omega_{max}}=0.71$ GeV.

Taking into account that  $\sigma(e^+e^-\to\phi\to\pi^0\rho, m_{\phi})<
182$ nb, for the cross section of the process
 $e^+e^-\to\phi\to\pi^0\rho\to\pi^0\eta\gamma$ in the region
 $\omega<100$ MeV one gets
\begin{equation}
\label{phieta}
\sigma(e^+e^-\to\phi\to\pi^0\rho\to\pi^0\eta\gamma, m_{\phi})
<6.37\cdot10^{-3}\ nb,
\end{equation}
wich corresponds to 
$BR(\phi\to\pi^0\rho^0\to\gamma\pi^0\eta,\omega<100\ MeV)
<1.5\cdot10^{-6}$.

So, the processes   $e^+e^-\to V^0\to\pi^0 V'^0\to\gamma\pi^0
\eta$ and  $e^+e^-\to V^0\to\eta V'^0\to\gamma\pi^0
\eta$ which are the  background to  
 $e^+e^-\to\gamma a_0\to\gamma\pi^0\eta$  are negligible
for  $BR(\phi\to\gamma a_0\to\gamma\pi^0\eta)$ greater than $10^{-5}$.

 As for the interference between
the  $e^+e^-\to\gamma a_0\to\gamma\pi^0\eta$ and  
$e^+e^-\to\phi\to\pi^0\rho\to\gamma\pi^0\eta$ processes, 
one can neglect it also due to the fact that  80\% of background 
contribution to the branching ratio under consideration is determined by the 
region $70<\omega<100$ MeV while the contribution of the primary 
process in this region is less than 20\% in the worst case when
the $a_0$ meson is relatively wide, see Sec.IV. 

Notice that if it should be possible to choose a cut on photon energy 
$\omega<50$ MeV the background would be one order of magnitude less
than previously stated while the primary processes 
$e^+e^-\to\phi\to\gamma f_0(a_0)\to\gamma\pi^0\pi^0(\eta)$ would not 
decrease considerably.

\section{ Mixing \lowercase{$f_0$} and \lowercase{$\sigma$} mesons.}
\subsection{\lowercase{$q^2\bar q^2$} and \lowercase{$q\bar q$} models.}

The background from processes
$e^+e^-\to V^0\to\pi^0 V'^0\to\gamma\pi^0\pi^0(\eta)$ described above are
negligible up to $BR(\phi\to\gamma\pi^0\pi^0(\eta), \omega<100\ MeV)=
5\cdot10^{-6}(10^{-5})$. 
In the meantime the one-loop calculation in the frame
of the chiral perturbation theory of $BR(\phi\to\gamma K\bar K\to\gamma
\pi\pi(\eta))$ at $\omega<100$ MeV leads to the number of $10^{-5}$ order
of magnitude, see \cite{chiral}. So, in such a theory the strong interference
effects with the reaction $\phi\to\gamma f_0(a_0)\to\gamma\pi\pi(\eta)$ are
predicted. But one cannot restrict oneself to the one-loop approximation
only in the region discussed  ($\sim1$ GeV), see, for example,
 \cite{sigma}. In view of the fact that all the  corrections to the amplitude
in the chiral model cannot be  taken into account, we treat them in the
 phenomenological way considering the scalar particle called $\sigma$ meson
that is strongly coupled with  $\pi\pi$ channel and, in view of mixing with
$f_0$ meson, can change considerably the photon energy  differential
 cross section. The parameters of the $\sigma$ meson we obtain
 from  fitting  the $\pi\pi$ scattering data.

Let us consider the reaction $e^+e^-\to\phi\to\gamma (f_0+\sigma)\to\gamma
\pi^0\pi^0$ with regard to the mixing of the $f_0$ and $\sigma$ mesons.
Below is the formalism in the frame of which we study this problem.

  We consider the one loop mechanism of the $R$ meson production, where
$R=f_0,\sigma$, through the charged kaon loop, $\phi\to K^+K^-\to\gamma R$,
see \cite{{achasov-89},{isgur-93},{molecule},{achasov-95}}. The symbolic
diagram is shown in Fig.3(a). The amplitude $\phi\to\gamma R$ in the rest
frame of the the $\phi$ meson is parametrized in the following manner
\begin{equation}
\label{main}
M=g(m)g_{RK^+K^-}\vec e(\phi)\vec e(\gamma)
\end{equation}
where $m^2=(k_1+k_2)^2=s-2\sqrt{s}\omega$ is the invariant mass of the
 $\pi\pi$ system, $\vec e(\phi)$ and $\vec e(\gamma)$ are the polarization
vectors of the $\phi$ meson and photon respectively.
The expression of the $g(m)$ function obtained in the four-quark
$(q^2\bar q^2)$ model is written down in
\cite{achasov-89}. \footnote{ For a convenience, in this paper,
we use a different parametrization of the amplitude (\ref{main}).
Please, compare with \cite{achasov-89,molecule,achasov-95,interference}}.

Notice that in the four-quark and two-quark models the scalar mesons
are considered to be point like objects \cite{{achasov-89},{achasov-95}}
but in the model of the $K\bar K$ molecule they are considered as  extended
ones \cite{weinstein-90}.

 Taking into account the mixing of the $f_0$ and $\sigma$ mesons,
the amplitude of the reaction
 $e^+(p_1)e^-(p_2)\to\phi\to\gamma (f_0+\sigma)\to\gamma(q)
\pi^0(k_1)\pi^0(k_2)$ is written  in the following way \cite{achasov-84,zphys}
\begin{equation}
M=e\bar u(-p_1)\gamma^{\mu}u(p_2)\frac{em_{\phi}^2}{f_{\phi}}
\frac{e^{i\delta_B}}{sD_{\phi}(s))}g(m)(e(\gamma)^{\mu}-q^{\mu}
\frac{e(\gamma)p}{pq})\sum_{RR'}(g_{RK^+K^-}G^{-1}_{RR'}(m)
g_{R'\pi^0\pi^0})  \label{amplituda}
\end{equation}
where $s=p^2=(p_1+p_2)^2$, $g_R(m)\sim(pq)\to0$ at $(pq)\to0$ and
 $\delta_B$ is the phase of the elastic background, see (\ref{amplppkk}).
The constant $f_{\phi}$ is related to the electron width of the vector meson
decay
\begin{equation}
\label{elwidth}
\Gamma(V\to e^+e^-,s)=\frac{4\pi\alpha^2}{3}(\frac{m_V^2}{f_V})^2
\frac{1}{s\sqrt{s}}
\end{equation}

For the differential cross section one gets
\begin{equation}
\label{signal}
\frac{d\sigma_{\phi}}{d\omega}=\frac{\alpha^2\omega}{8\pi s^2}
\left(\frac{m_{\phi}^2}{f_{\phi}}\right)^2\frac{|g(m)|^2}{|D_{\phi}(s)|^2}
\sqrt{1-\frac{4m_{\pi}^2}{m^2}}(c+\frac{c^3}{3})
\left|\sum_{RR'}(g_{RK^+K^-}G^{-1}_{RR'}g_{R'\pi^0\pi^0})\right|^2
H(s,\omega_{min})
\end{equation}
where $\omega=|\vec q|$ is the energy of the photon, $c$ is the cut on
$\cos\theta_{\gamma}$, where $\theta_{\gamma}$ is the angle between
the photon momentum direction and the beam in
the center of mass frame of the considered reaction:
$-c\leq\cos\theta_{\gamma}\leq c$. The function $H(s,\omega_{min})$
takes into account the radiative corrections, the contribution of which
is about 20\% \cite{fadin}.
\begin{equation}
H(s,\omega_{min})=\frac{1}{|1-\Pi(s)|^2}\{1+
\frac{2\alpha}{\pi}[(L-1)\ln\frac{2
\omega_{min}}{\sqrt{s}}+\frac{3}{4}L+\frac{\pi^2}{6}-1]\}
\end{equation}
where $\omega_{min}$ is the minimal photon energy registered in the
experiment, $L=\ln\frac{s}{m_e^2}$ is the "main" logarithm.
The electron vacuum polarization to the one order of $\alpha$ is
\begin{equation}
\Pi(s)=\frac{\alpha}{3\pi}(L-\frac{5}{3}),
\end{equation}
the contribution of the muon and light hadron has been omitted.
Describing the photon spectrum, we shall use the photon energy cuts
$20<\omega<100$ MeV that allows the separation of the signal from the other 
states contributing in the differential cross section. But,
 calculating the branching ratios of the processes,
 we shall neglect the $\omega_{min}$,
as it was done above, and use the photon energy cut  $\omega<100$ MeV.

The matrix of the inverse propagator has the form
\begin{displaymath}
{G_{RR'}(m)}=
\left( \begin{array}{cc}
D_{f_0}(m)&-\Pi_{f_0\sigma}(m)\\
-\Pi_{\sigma f_0}(m)&D_{\sigma}(m)
\end{array} \right)
\end{displaymath}

For the propagator of the scalar particle we use the following expression
\begin{equation}
\label{propagator}
D_R(m)=m_R^2-m^2+\sum_{ab}g_{Rab}[Re P_R^{ab}(m_R^2)-P_R^{ab}(m)],
\end{equation}
where $\sum_{ab}g_{Rab}[Re P_R^{ab}(m_R^2)-P_R^{ab}(m)]=\Pi_R(m)=
\Pi_{RR}(m)$ takes into account
the finite width corrections of the resonance which are the one loop
contribution to the self-energy of the $R$ resonance from the two-particle
intermediate  $ab$ states. In the  $q^2\bar q^2$ model of the scalar particle
and in the model of the $K\bar K$ molecule the $f_0$ and $a_0$ mesons
are strongly coupled with the $K\bar K$ channel, since they are just under
the threshold of this channel.
 The ordinary resonance expression of the propagator, in view of this
coupling, is changed considerably and the account of
$\sum_{ab}g_{Rab}[Re P_R^{ab}(m_R^2)-P_R^{ab}(m)]$
  corrections is necessary, see
\cite{achasov-84,achasov-89,molecule,inadequacy}. Notice that the expression
(\ref{signal}) for mixing of the $f_0$ and $\sigma$ mesons takes into account
 the all order of the perturbation theory, the symbolic diagrams are shown
 in Fig.3.

For the pseudoscalar $ab$ mesons and $m_a\geq m_b,\ m^2>m_+^2$ one has
\begin{equation}
\label{polarisator}
P^{ab}_R(m)=\frac{g_{Rab}}{16\pi}\left[\frac{m_+m_-}{\pi m^2}\ln
\frac{m_b}{m_a}+\rho_{ab}\left(i+\frac{1}{\pi}\ln\frac{\sqrt{m^2-m_-^2}-
\sqrt{m^2-m_+^2}}{\sqrt{m^2-m_-^2}+\sqrt{m^2-m_+^2}}\right)\right]
\end{equation}
In other regions of $m$ one can obtain the $P_R^{ab}(m)$ by analytical
continuation of Eq.(\ref{polarisator}),
 see \cite{achasov-95,interference,ach-81}.

The constants  $g_{Rab}$ are related to the width
\begin{equation}
\Gamma(R\to ab,m)=\frac{g_{Rab}^2}{16\pi m}\rho_{ab}(m),
\label{f0pipi}
\end{equation}
where $\rho_{ab}(m)=\sqrt{(m^2-m_+^2)(m^2-m_-^2)}/m^2$ and
 $m_{\pm}=m_a\pm m_b$.

Nondiagonal elements of the matrix $G_{RR'}(m)$ are the transitions caused by
the resonance mixing due to the final state interaction which occurred
 in the same decay channels $R\to (ab)\to R'$. We write them down in the
following manner \cite{achasov-84,zphys}
\begin{equation}
\Pi_{RR'}(m)=\sum_{ab}g_{R'ab}P_R^{ab}(m)+C_{RR'},
\end{equation}
where the constants $C_{RR'}$ take into account effectively the contribution
of $VV,\ 4P$ and other intermediate states and incorporate the
subtraction constants for the  $R\to(PP)\to R'$ transitions. In the
four-quark and two-quark models we treat these constants as a free parameters.

\subsection{ Model of $K\bar K$ molecule.}

The model of the $K\bar K$ molecule was formulated in \cite{weinstein-90}
and developed  in the papers \cite{{isgur-93},{molecule}}, as applied to the
$\phi\to\gamma f_0(a_0)\to\gamma\pi\pi(\eta)$ decay.

A specific feature of the molecular model is the narrow structure of the
$f_0$ and $a_0$ mesons.

Really, the width of the weakly bound, quasi-stable system cannot be larger
than the bound energy which is $\epsilon=10-20$ MeV for the $K\bar K$ mesons,
see, for example, \cite{weinstein-90}. So, in the molecular model the
effective width of the $f_0$ and $a_0$ mesons is $\Gamma_{eff}< 20$ MeV.
The $\pi\pi$ scattering data, as we shall see below, permit the width of
the $f_0$ meson to be $\Gamma_{eff}=0.01-0.03$ GeV. Such widths, with some
stretch of the  interpretation \footnote{
Strictly speaking, the width should be much less then the bound energy.}
 can be associated with the $K\bar K$ bound state.
But, nevertheless, we suppose that the latest experimental data
are difficult ( apparently impossible ) to understand  in the framework of
this model as they point to the rather wide  $f_0$ resonance with the
 $\Gamma_{eff}\sim 40-100$ MeV, see \cite{{pdg-96},{gams},{dzirba},{brabson}}.
The objections against the  $K\bar K$ bound state interpretation were also
presented in \cite{pennington}. As for the  $a_0$ resonance,
 it seemed to be always too wide for a such interpretation,
 $\Gamma_{eff}\sim 50-100$ MeV, see
\cite{pdg-84}.  The latest data confirm this. The $a_0$ meson with
the  $\Gamma_{eff}\simeq90\pm11$ MeV was observed in the reaction
 $\pi^-p\to\pi^0\eta n$ (Brookhaven), see \cite{dzirba}.

In the model of the scalar  $K\bar K$ molecule \cite{isgur-93} the function
$g(m)$was calculated in \cite{molecule}. In this paper it was found
that  the imaginary part of the $\phi\to\gamma f_0$ amplitude gives
$90\%$ of  $BR(\phi\to\gamma f_0\to\gamma\pi\pi)$.
In view of this, we take into account only the imaginary part of the $g(m)$.
In the transitions caused by the resonance mixing due to the final state
interaction only real intermediate states are involved ( the virtual states
are suppressed for an extended molecule ). Because of virtual states
suppression, we write in the model of  $K\bar K$ molecule the
 nondiagonal elements of the inverse propagator matrix in the following way
\begin{equation}
\Pi_{RR'}(m)=Im\sum_{ab}g_{R'ab}P_R^{ab}(m).
\end{equation}

As for the propagator of the $f_0$ meson, let us take the generally accepted
Breit - Wigner formulae.

   When $m>2m_{K^+}\  2m_{K^0}$\ ,
\begin{eqnarray}
\label{breit}
& & D_{f_0}(m)=M_{f_0}^2-m^2-i(\Gamma_0(m)+\Gamma_{K\bar K}(m))m
\nonumber\\
& &\Gamma_{K\bar K}(m)=\frac{g^2_{f_0K^+K^-}}{16\pi}(\sqrt{1-4m^2_{K^+}/m^2}+
\sqrt{1-4m^2_{K^0}/m^2})\frac{1}{m}\ .
\end{eqnarray}
As a parameter of the model we use the decay width
$\Gamma(f_0(a_0)\to\pi\pi(\eta))=\Gamma_0$ which is defined in
Eq.(\ref{f0pipi}). In other areas of $m$ one can obtain the
$D_{f_0}(m)$ by analytical continuation of Eq.(\ref{breit}),
see \cite{molecule,inadequacy}.

 Since the scalar resonances lie under the $K\bar K$ thresholds the position
of the peak in the cross section or in the mass spectrum does not coincide
with $M_{f_0}$, as one can see from analytical continuation of
Eq.(\ref{breit}) under the $K\bar K$ threshould \cite{molecule,inadequacy}.
 That is why it is necessary to renormalize the mass in the
 Breit - Wigner formulae
\begin{eqnarray}
M^2_{f_0} = m^2_R - \frac{g^2_{f_0K^+K^-}}{16\pi}(\sqrt{4m^2_{K^+}/
m_{f_0}^2-1}+\sqrt{4m^2_{K^0}/m_{f_0}^2-1})\ ,
\end{eqnarray}
where $m^2_{f_0}$ is the physical mass squared ( $m_{a_0} = 980$ MeV and
$m_{f_0}= 980$ MeV) while $M^2_{f_0}$ is the bare mass squared. So, the
physical mass is heavier than the bare one. This circumstance is especially
 important in the case of a strong coupling of the scalar mesons with
the $K\bar K$ channel as in the molecular models.

In the molecular model the coupling constant of the $f_0$ meson with the
$K\bar K$ channel is \cite{isgur-93}
\begin{equation}
\frac{g^2_{f_0K^+K^-}}{4\pi}=0.6\ GeV^2.
\end{equation}
Notice that in this model $m_{f_0}-M_{f_0}=24(10)$ MeV for
 $m_{f_0}=980(2m_{K^+})$. The value of $\Gamma_0$ is fixed in the region
$0.05-0.1$ GeV that leads, on average, to the effective widths
$\Gamma_{eff}\simeq0.01-0.03$ GeV.

\subsection{ Analyses of data on the $\pi\pi\to\pi\pi$ scattering.}

To fit the $\pi\pi$ scattering data we write the $s$-wave amplitude
of the $\pi\pi\to\pi\pi$ reaction with $I=0$ as the sum of the inelastic
resonance amplitude $T^{res}_{\pi\pi}$, in which the contributions of the
$f_0$ and $\sigma$ mesons  are taken onto account, and the amplitude of
the elastic background \cite{achasov-84,zphys}
\begin{equation}
\label{eqpipi}
T(\pi\pi\to\pi\pi)=\frac{\eta^0_0e^{2i\delta^0_0}-1}{2i\rho_{\pi\pi}}=
\frac{e^{2i\delta_B}-1}{2i\rho_{\pi\pi}}+e^{2i\delta_B}T^{res}_{\pi\pi},
\end{equation}
where
\begin{equation}
T^{res}_{\pi\pi}=\sum_{RR'}\frac{g_{R\pi\pi}g_{R'\pi\pi}}{16\pi}
G^{-1}_{RR'}(m)
\end{equation}
The elastic background phase $\delta_B$ is taken in the form
$\delta_B=\theta\rho_{\pi\pi}(m)$, where  $\theta\sim60^{\circ}$.
The expressions like Eq.(\ref{eqpipi}) are commonly used
for the fitting in the wide energy
region. This fairly simple approximation works to advantage
in the $f_0$ resonance region both in the $\pi\pi\to\pi\pi$ and in the
$\pi\pi\to K\bar K$ reactions. The $\pi\pi\to K\bar K$ channel is the first
important inelastic channel of the $\pi\pi$ scattering in the $s$-wave.
 Defying the relations
\begin{eqnarray}
&&g_{f_0K\bar K}=\sqrt{2}g_{f_0K^+K^-}=\sqrt{2}g_{f_0K^0\bar K^0} \nonumber\\
&&g_{\sigma K\bar K}=\sqrt{2}g_{\sigma K^+K^-}=\sqrt{2}g_{\sigma K^0\bar K^0},
\end{eqnarray}
we get for inelasticity the following expression
\begin{eqnarray}
&&\eta^0_0=1; \qquad m<2m_{K^+} \nonumber\\
&&\eta^0_0=\sqrt{1-2\rho_{\pi\pi}(m)\rho_{K^+K^-}(m)|T_{K\bar K}|^2};\qquad
2m_{K^+}<m<2m_{K^0} \nonumber\\
&&\eta^0_0=\sqrt{1-2\rho_{\pi\pi}(m)(\rho_{K^+K^-}(m)+\rho_{K^0\bar K^0}(m))
|T_{K\bar K}|^2};\qquad m>2m_{K^0}
\end{eqnarray}
where the amplitude of the $\pi\pi\to K\bar K$ process is
\begin{equation}
\label{amplppkk}
T_{K\bar K}=e^{i\delta_B}\sum_{RR'}\frac{g_{R\pi\pi}g_{R'K\bar K}}{16\pi}
G^{-1}_{RR'}(m).
\end{equation}

\section{ Results and discussions.}

\subsection{\lowercase{$f_0(980)$} resonance.}

In the four-quark model  we consider the following parameters as  free:
the coupling constant of the $f_0$ meson to the $K\bar K$
channel $g_{f_0K^+K^-}$, the coupling constant of the $\sigma$ meson
to the $\pi\pi$ channel $g_{\sigma\pi\pi}$, the constant $C_{f_0\sigma}$,
the ratio $R=g^2_{f_0K^+K^-}/g^2_{f_0\pi^+\pi^-}$, the phase $\theta$ of the
elastic background and the $\sigma$ meson mass. The mass of the $f_0$ meson
is restricted to the region $0.97<m_{f_0}<0.99$ GeV.  We treat the  $\sigma$
meson as an ordinary two-quark state  the coupling constant of which with the
$K\bar K$ channel must be related in the naive quark model to the constant
$g_{\sigma\pi\pi}$. One gets $g_{\sigma K^+K^-}=g_{\sigma\pi^+\pi^-}/2$.
From the analyses of the radiative decays of the $J/\psi$ meson it is evident
that the strange quark production is suppressed by a factor of
$\lambda\simeq1/2$ in comparison with $u$ and $d$ quarks \cite{anisovich}
\begin{equation}
u\bar u:d\bar d:s\bar s=1:1:\lambda
\end{equation}
This leads to the additional suppression of the coupling constant of the
$\sigma$ meson with the $K\bar K$ channel, $g_{\sigma K^+K^-}=\sqrt{\lambda}
 g_{\sigma\pi^+\pi^-}/2\simeq0.35g_{\sigma\pi^+\pi^-}$. As it was found, the
fitting  depends  on the constant  $g_{\sigma K^+K^-}$ very weakly and
we drop this constant. We do not consider the coupling
of the $\sigma$ meson with the $\eta\eta$ channel as well. In the naive
quark model this coupling is relatively suppressed. If the angle of the
 $\eta-\eta'$ mixing is chosen to be $\theta_{\eta\eta'}=-18^{\circ}$ one
gets $g_{\sigma\eta\eta}\simeq(\sqrt{2}/3)g_{\sigma\pi^+\pi^-}$ .

We use the  data presented in the papers \cite{{hyams},{estabrook}}
for the fitting of the phase $\delta^0_0$ of the $\pi\pi$ scattering and
the data presented in the papers \cite{{hyams},{martin}}
 for fitting the inelasticy $\eta^0_0$.

The fitting shows that in the four quark model a number of parameters
describe well enough the $\pi\pi$ scattering in the region $0.7<m<1.8$ GeV.
 Our results in the $(q^2\bar q^2)$ model are tabulated in the Table I.
 In this table the following parameters and predictions are listed:
 the effective widths of the
$f_0$ meson $\Gamma_{eff}$, the branching ratios $BR_{f_0+\sigma}(BR_{f_0})=
3\cdot BR(\phi\to\gamma(f_0+\sigma)\to\gamma\pi^0\pi^0)(3\cdot
BR(\phi\to\gamma f_0\to\gamma\pi^0\pi^0))$ of the decays
$\phi\to\gamma (f_0+\sigma)\to\gamma\pi\pi$ and  $\phi\to\gamma f_0
\to\gamma\pi\pi$   integrated over the range $0.7<m<m_{\phi}$
 (  corresponding to $\omega<250$ MeV ) and
 taking into account the energy cut  $\omega<100$ GeV ( the range
 $0.9<m<m_{\phi}$ )\footnote{ Notice that $BR(\phi\to\gamma \pi^0\pi^0)=
 (1/3)BR(\phi\to\gamma \pi\pi)$.}. Besides that we study the dependence
of the fitting parameters on the mass of the $f_0$ meson. All dimensional
quantities in the tables are presented in the units of GeV  or in $GeV^2$.
 The picture for one set of the parameters is shown in Fig 4.

As it is seen from the Table I, one can change the effective width of the
$f_0$ meson from the very narrow $\Gamma_{eff}\simeq10$ MeV to the relatively
wide $\Gamma_{eff}\simeq100$ MeV, varying the parameters in the four-quark
model. The parameters of the  $\sigma$ meson are
approximately unchanged  $m_{\sigma}\simeq1.48$ GeV, $g^2_{\sigma\pi\pi}/
4\pi\simeq1.8\ GeV^2$. The decay width of the $\sigma$ meson to $2\pi$ is
$\Gamma_{2\pi}(m_{\sigma})\simeq300$ MeV. Besides, it is seen that in
spite of the $f_0$ and $\sigma$ mixing  the branching ratios
of the decays
 $\phi\to\gamma f_0\to\gamma\pi\pi$ and $\phi\to\gamma
(f_0+\sigma)\to\gamma\pi\pi$ are approximately the same and have the order of
$10^{-4}$.

Recall that in the four-quark model the coupling of the $a_0$ and $f_0$
states to the $K\bar K$ channel is strong, superallowed by OZI rule
\cite{{jaffe-77},{achasov-84}}, $g^2_{f_0(a_0)}/4\pi\gtrsim1\ GeV^2$.
Besides, in the $q^2\bar q^2$ model, where the $f_0$ meson has a symbolic
structure $f_0=s\bar s(u\bar u+d\bar d)/\sqrt{2}$, the coupling of the $f_0$
meson to the $\pi\pi$ channel is relatively suppressed. In this case one
should consider $R=1,2$ as a relative enhancement since the comparison
 should be made  with $R=g^2_{f_0K^+K^-}/g^2_{f_0\pi^+\pi^-}=\lambda/4=1/8$
in the case of $f_0=(u\bar u+d\bar d)/\sqrt{2}$.

The three last lines in Table I are devoted to the $s\bar s$ model of
the $f_0$ meson. In such a model the $f_0$ meson is considered as a
point-like object, i.e. in the  $K\bar K$ loop, $\phi\to K^+K^-\to\gamma
f_0$ and in the transitions caused by the resonance mixing we consider
both the real and the virtual intermediate states. It is typical for
the $s\bar s$ model of the $f_0$ meson that the coupling with the $\pi\pi$
channel is relatively suppressed. In this sense this model is different from
the $q^2\bar q^2$ model by the small coupling constant $g_{f_0K^+K^-}$ only,
in the $s\bar s$ model $g^2_{f_0K^+K^-}/4\pi\sim0.3\ GeV^2$.
The small coupling constant $g_{f_0K^+K^-}$ arises from the requirement
of the relation between constants $g_{a_0\pi\eta}$ and $g_{f_0K^+K^-}$ in
the two-quark model. If the angle of the  $\eta-\eta'$ mixing is chosen
to be $\theta_{\eta\eta'}=-18^{\circ}$ one gets $g_{a_0\pi\eta}=
2\cos(\theta_{\eta\eta'}+\theta_p)g_{a_0K^+K^-}=1.6g_{a_0K^+K^-}$, where
 $\theta_p=54.3^{\circ}$ is the "ideal"  mixing angle. Taking into account
that in the $s\bar s$ model the $g_{f_0K^+K^-}=\sqrt{2}g_{a_0K^+K^-}$ one
gets $g_{a_0\pi\eta}=1.13g_{f_0K^+K^-}$. The requirement, that the
$\Gamma(a_0\to\pi\eta)<0.1$ GeV, leads to the constraint
$g^2_{f_0K^+K^-}/4\pi<0.5\ GeV^2$. As it is seen from the Table I,
in the $s\bar s$ model of the $f_0$ meson the branching ratio
$BR(\phi\to\gamma (f_0+\sigma)\to\gamma\pi\pi)\simeq5\cdot10^{-5}$.

Notice again that the structure $f_0=(u\bar u+d\bar d)/\sqrt{2}$ with
$R=1/4-1/8$ is eliminated completely by the data on the $\pi\pi$ scattering.
The data on the $\pi\pi$ scattering permit the ratio
 $R=g^2_{f_0K^+K^-}/g^2_{f_0\pi^+\pi^-}$ to range over the relatively large
 interval: $R=1-20$. This interval is somewhat larger than expected
 previously, see
\cite{achasov-84,achasov-89,achasov-95}. This is due to the fact that
previous experimental data pointed in favour of the narrow
structure of the $f_0$ meson $\Gamma_{eff}\simeq20-50$ MeV \cite{pdg-84}.
But now, in the decays of the $J/\psi$ meson, in the reaction
 $J/\psi\to\phi f_0$ \cite{falvard}, the $f_0(980)$-peak with
 $\Gamma_{eff}\simeq80$ MeV is observed. The recent GAMS data on the
reaction $\pi^-p\to\pi^0\pi^0n$ \cite{gams} at  large momentum transfer
show the $f_0(980)$-peaks with  $\Gamma_{eff}\simeq50$ MeV .
In this reaction the E 852 Collaboration \cite{brabson}
 also observes the $f_0(980)$-peak with $\Gamma_{eff}=61\pm18$ MeV
at large momentum transfer.

In view of this, fitting the data on the $\pi\pi$ scattering, we considered
also  relatively small $R$ ($R\simeq1-5$) which leads to
the  $\Gamma_{eff}\lesssim100$ MeV.

The parameter set for the model of the $K\bar K$ molecule is tabulated
in  Table II and the illustrative graphics are shown in the Fig.5.

As  seen in Table II, the data on the  $\pi\pi$ scattering in the
model of the $K\bar K$ molecule permit $\Gamma_{eff}=10-30$ MeV which is
likely for the bound  $K\bar K$ state. Those values of the
effective widths could be achieved with  different parameters of the
$K\bar K$ model. In this case, the branching ratios  of the decays
 $\phi\to\gamma f_0\to\gamma\pi\pi$ and
$\phi\to\gamma (f_0+\sigma)\to\gamma\pi\pi$ remain approximately
the same and have the order of $10^{-5}$.

Notice that the parameters obtained for the $\sigma$ meson are in  good
agreement with those discussed in the literature for the
  $\epsilon(1300)$ meson
\cite{pdg-84}, currently known as  $f_0(1300)$ \cite{particle-94} or
 $f_0(1370)$ \cite{pdg-96} meson. Among all decay channels of this particle
we have considered the main $\pi\pi$ only, $\Gamma_{2\pi}(m_{\sigma})\simeq
300$ MeV. The channels $K\bar K$ and
 $\eta\eta$, as it was found, are relatively suppressed and change
 the parameters of the
fitting very weakly. The influence of the $4\pi$ channel we  analyzed
separately and found that the account of it does not change our results
essentially.

The decay width of the  $\sigma$ meson to $4\pi$ was approximated by the
polynomial $m\Gamma_{4\pi}(m)=0.32\cdot(m-0.7)(m-0.6)$ so that
$\Gamma_{4\pi}(m_{\sigma}=1.5)=150$ MeV and $\Gamma_{4\pi}(m=0.7)=0$.
The analyses of the decay  $f_0(1520)\to4\pi$
shows that a such approximation is rather resonable, see \cite{f01520}.
The account of $im\Gamma_{4\pi}(m)$ in the propagator of the $\sigma$ meson
leads to the increasing of the coupling constant $g_{\sigma\pi\pi}$  and
the $\sigma$ meson become correspondingly wider $\Gamma_{\sigma}(m_{\sigma})=
\Gamma_{2\pi}(m_{\sigma})+\Gamma_{4\pi}(m_{\sigma})\simeq550$ MeV. The
branching ratios of the decays  $\phi\to\gamma f_0\to\gamma\pi\pi$ and
$\phi\to\gamma (f_0+\sigma)\to\gamma\pi\pi$ are changed by 8\%
and our main results, after taking into account the
$4\pi$ channel, remain unchanged .

It is seen from the Tables I and II that the regions of the constants
$g_{f_0K^+K^-}$ and $g_{f_0\pi^+\pi^-}$ in the four-quark model and in the
model of the $K\bar K$ molecule overlap. It does not mean,
of course, that in this case the models are the same since in the model of the
$K\bar K$ molecule we consider the real intermidiate states only in the
the transitions caused by the resonance mixing due to the final state
interaction. The suppression of the virtual states is typical
for an extended molecule. At the same time, in the $q^2\bar q^2$ model
we consider both the real and the virtual intermidiate states.

\subsection{\lowercase{$a_0(980)$} resonance.}

As for the reaction $e^+e^-\to\gamma\pi^0\eta$  the similar analyisis
of the $\pi\eta$ scattering cannot be performed directly. But, our analysis
of the final state  interaction for the $f_0$ meson production show that
the situation does not changed radically, in any case in the region
$\omega<100$ MeV. Hence, one can hope that the final state interaction
in the  $e^+e^-\to\gamma a_0\to\gamma\pi\eta$ reaction will not strongly
affect the predictions in the region $\omega<100$ MeV.

Recall that in the four-quark model  the $a_0$ meson has a symbolic structure
$a_0=s\bar s(u\bar u-d\bar d)/\sqrt{2}$ and the following relations are valid
\cite{achasov-84,achasov-89}:
\begin{eqnarray}
\label{4-qq}
&&g_{a_0K^0\bar K^0}=-g_{a_0K^+K^-};\qquad g_{a_0K^+K^-}=g_{f_0K^+K^-}
\nonumber \\
&&g_{a_0\pi\eta}=\sqrt{2}g_{a_0K^+K^-}\sin(\theta_p+\theta_{\eta\eta'})=
0.85g_{a_0K^+K^-} \nonumber \\
&&g_{a_0\pi\eta'}=-\sqrt{2}g_{a_0K^+K^-}\cos(\theta_p+\theta_{\eta\eta'})=
-1.13g_{a_0K^+K^-}.
\end{eqnarray}
Based on the analysis of $\pi\pi$ scattering with regard to
Eq.(\ref{4-qq}) we predict the quantities of the
 $BR(\phi\to\gamma a_0\to\gamma\pi\eta)$ in the  $q^2\bar q^2$ model.
The results are shown in the Table III.

In the propagator of the $a_0$ meson we consider the final width corrections
corresponding the $\pi\eta,K\bar K, \pi\eta'$ channels, see
Eq.(\ref{polarisator}). As is seen from Table III, in the  $q^2\bar q^2$ model
of the $a_0$ meson the
$BR(\phi\to\gamma a_0\to\gamma\pi\eta)\simeq\cdot10^{-4}$.

The last two lines in  Table III relate to the $s\bar s$ model
of the $f_0$ meson in which the $a_0$ meson is treated as a two-quark
$(u\bar u-d\bar d)/\sqrt{2}$ state. In this model \cite{achasov-84,achasov-89}
\begin{eqnarray}
&&g_{f_0K^+K^-}=\sqrt{2}g_{a_0K^+K^-};\qquad g_{a_0K^0\bar K^0}=-g_{a_0K^+K^-}
\nonumber \\
&&g_{a_0\pi\eta}=2g_{a_0K^+K^-}\cos(\theta_p+\theta_{\eta\eta'})=
1.6g_{a_0K^+K^-} \nonumber \\
&&g_{a_0\pi\eta'}=2g_{a_0K^+K^-}\sin(\theta_p+\theta_{\eta\eta'})=
1.2g_{a_0K^+K^-}.
\end{eqnarray}
Recall that in this case the $\phi\to\gamma a_0\to\pi\eta$ decay
is suppressed by the OZI rule, as discussed in the  Introduction.
The OZI suppression means that the real part
of the $K^+K^-$ loop is compensated by the virtual intermediate states
$K^{*+}K^{-}$, $K^{+}K^{*-}$, $K^{*+}K^{*-}$ and so on.
 Because there is a real two-particle intermediate  $K^+K^-$ state the
  OZI rule violating imaginary part of the decay amplitude is relatively
large \cite{achasov-89}. In view of this, we consider  the imaginary part
 of the $\phi\to\gamma a_0(q\bar q)$ decay amplitude only.
 As one can see from Table III in the two-quark model of the $a_0$ meson the
$BR(\phi\to\gamma a_0\to\gamma\pi\eta)\simeq8\cdot10^{-6}$.

In the model of the $K\bar K$ molecule the relations
$g_{a_0K^0\bar K^0}=-g_{a_0K^+K^-}$ and $g_{a_0K^+K^-}=g_{f_0K^+K^-}$ are
valid, in this case $BR(\phi\to\gamma a_0\to\gamma\pi\eta)\sim10^{-5}$.
The results are listed in Table IV.

\subsection{ Metamorphosis of the \lowercase{$a_0(980)$} and
 \lowercase{$f_0(980)$} resonances.}

In this section we would like  to discuss the definition
of the scalar resonance effective widths listed in the tables.

According to the current view, away from the thresholds of the reactions
the production amplitude of the scalar resonance  has the form
$A=g_ig_j/(s-m_R^2+im_R\Gamma)$, where the quantities $g_i,g_j$ are the
constants. So, the form of the resonance defined by Breit-Wigner formulae
has a universal form and is described by the two parameters: the mass and
the width.  Most resonances, especially narrow ones,  satisfy this
condition of universality and are adequately described by the
Breit-Wigner form. But, such a simple situation is not always true.

The availability of the channels in the immediate vicinity of the resonance
point, with which the resonance strongly couples, leads to energy-dependent
$g_i,g_j$ quantities. Also, the denominator becomes an intricate
function of the energy, see Eq.(\ref{propagator}), Eq.(\ref{polarisator}),
in which both the real and imaginary parts vary rapidly. This leads
to a considerable distortion of the simple Breit-Wigner formulae, see
\cite{{achasov-84},{ach-81},{achasov-gamma}}.
The property of the universality is lost in this case, the resonance form,
in particular the visible width,  dependeds on the specific
production mode of the resonance.

This phenomena can be brightly demonstrated by the example of the
scalar $a_0$ and $f_0$ resonances.

In the peripheral production of the $a_0$ meson, for example, in the reaction
 $\pi^- p\to\pi^0\eta n$, the mass spectrum of the $\pi\eta$ system is
determined by the expression
\begin{equation}
\label{spectr}
\frac{dN_{\pi\eta}}{dm}\sim\frac{m^2\Gamma_{a_0\pi\eta}(m)}{|D_{a_0}(m)|^2},
\end{equation}
where $m$ is the mass of the $\pi\eta$ system. In Table III the effective
widths  corresponding to this distribution are listed ( in the four-quark
model $g_{a_0\pi\eta}=0.85g_{a_0K^-K^+}$ ). In the $e^+e^-\to\phi
\to\gamma a_0\to\gamma\pi\eta$ reaction the mass spectrum of the
 $\pi\eta$ system is determined by the expression
 \begin{equation}
\label{signalfora0}
\frac{d\sigma_{\phi}}{dm}=\frac{\alpha^2m}{16\pi s^3}
\left(\frac{m_{\phi}^2}{f_{\phi}}\right)^2\frac{|g(m)|^2}{|D_{\phi}(s)|^2}
(s-m^2)\sqrt{1-\frac{4m_{\pi}^2}{m^2}}(c+\frac{c^3}{3})
\left|\frac{g_{a_0K^+K^-}g_{a_0\pi^0\eta}}{D_{a_0}(m)}\right|^2
H(s,\omega_{min}).
\end{equation}
 For the constant listed in the Table III, at $g_{a_0K^+K^-}^2/4\pi>1\ GeV^2$
 in the  $q^2\bar q^2$ model, the visible width of the mass spectrum of
 the $\pi\eta$ system is twice as large as the effective one determined
 by Eq.(\ref{spectr}). For illustration those two spectrum
 are depicted together in Fig.6(a).

So,  considering the specific resonance production mechanism
(the one charge kaon loop production ) widens the spectrum twice as much.

Notice that such an effect is especially strong  at the large coupling
constants, i.e. in the four-quark model. With the coupling constant
decrease the effect dies out, for example, at the
$g_{a_0K^+K^-}^2/4\pi=0.72\ GeV^2$, as it is seen in Fig.6(b), the width of
the mass spectrum of the $\pi\eta$ system is one and a half as large as
 the one determined by Eq.(\ref{spectr}), and at the
 $g_{a_0K^+K^-}^2/4\pi=0.3\ GeV^2$ this effect is negligible.

For the  $f_0$ resonance the situation is somewhat the same with the
difference that the width of the resonance depends on the free
parameter $R$. At the $R=1-3$ the resonance is relatively wide and
the distinction between the widths of the $\pi\pi$ mass spectra in the
$e^+e^-\to\phi\to\gamma\pi\pi$ reaction and in the distribution
\begin{equation}
\label{spectrpipi}
\frac{dN_{\pi\pi}}{dm}\sim\frac{m^2\Gamma_{f_0\pi\pi}(m)}{|D_{f_0}(m)|^2},
\end{equation}
 is strong, see Fig.7(a). The phase and the inelasticity  are shown in Fig.8
 in this case .
  At large $R$, when the effective resonance
width $\Gamma_{eff}\sim30$ MeV, the difference is not so noticeable.
The illustrative picture is shown in the Fig.7(b).

Notice that when analyzing the definite processes it is needed  to
take into account the initial and final state interactions.
 Let us consider the
 $f_0$ production in the $J/\psi$ decay, where the peak with the visible
width $\Gamma\simeq80$ MeV is observed \cite{falvard}. The spectrum of the
 $\pi\pi$ system is determined by the expression
\begin{equation}
\label{spectrjpsipipi}
\frac{dN_{\pi\pi}}{dm}\sim\frac{m^2\Gamma_{f_0\pi\pi}(m)}{|D_{f_0}(m)|^2}
\cdot\left|\frac{ D_{\sigma}(m)+(1+\chi)\Pi_{f_0\sigma}(m)+
\chi(g_{\sigma\pi\pi}/g_{f_0\pi\pi})D_{f_0}(m)}
{D_{\sigma}(m)-\Pi_{f_0\sigma}^2/D_{f_0}(m)}\right|^2,
\end{equation}
where $\chi$ is a relative weight of the $\sigma$ meson production.
For the width of Eq.(\ref{spectrjpsipipi}) to be the visible width
 $\Gamma\simeq80$ MeV one needs to take not only the effective width
$\Gamma_{eff}\simeq80$ MeV, but to consider $\chi=1$ as well.
If one considers  that the $\sigma$ meson is not produced in the initial
state, i.e. $\chi=0$, but is produced as a result of the mixing with
the $f_0$ meson only in the final state, then the visible width of the
spectrum Eq.(\ref{spectrjpsipipi}) is equal to $\Gamma\simeq40$ MeV.

\section{Conclusion.}

We have shown that the vector meson contribution
$e^+e^-\to V^0\to\pi^0 V'^0\to\gamma\pi^0\pi^0(\eta)$ to the
 $e^+e^-\to\gamma\pi^0\pi^0(\eta)$ reaction is negligible in comparison with
 the scalar   $f_0(a_0)$ meson one
 $e^+e^-\to\phi\to\gamma f_0(a_0)\to\gamma\pi^0\pi^0(\eta)$ for
 $BR(\phi\to\gamma f_0(a_0))$ greater than $5\cdot10^{-6}(10^{-5})$,
in the photon energy region less than 100 MeV.

Based on the two-channel analysis of the $\pi\pi$ scattering we predict
for the  $\phi\to\gamma (f_0+\sigma)\to\gamma\pi\pi$ reaction
in the region $0.7<m<m_{\phi}$ that the
$BR(\phi\to\gamma (f_0+\sigma)\to\gamma\pi\pi)\sim10^{-4}$ and
$BR(\phi\to\gamma a_0\to\gamma\pi\eta)\sim10^{-4}$ in the $q^2\bar q^2$
model.

In the model of the $K\bar K$ molecule we get
$BR(\phi\to\gamma (f_0+\sigma)\to\gamma\pi\pi)\sim10^{-5}$ and
$BR(\phi\to\gamma a_0\to\gamma\pi\eta)\sim10^{-5}$.

In the two-quark $s\bar s$ model of the $f_0$ meson we obtain
$BR(\phi\to\gamma (f_0+\sigma)\to\gamma\pi\pi)\simeq5\cdot10^{-5}$ and
taking into account the imaginary part of the decay amplitude only,
as the main one, we get
$BR(\phi\to\gamma a_0(q\bar q)\to\gamma\pi\eta)\simeq8\cdot10^{-6}$.

Notice that the variants listed in the Table I describe equally well the
$\pi\pi$ scattering data, see, for example, Fig.4 and Fig.8.
 We could not find the parameters at which
the data on the $\pi\pi$ scattering are described well but the
$BR(\phi\to\gamma (f_0+\sigma)\to\gamma\pi\pi)$ is less then $10^{-5}$.
Or, more precisely, we could get  $BR(\phi\to\gamma (f_0+\sigma)
\to\gamma\pi\pi)\simeq10^{-5}$, only if we droped out the real part of the
$K^+K^-$ loop in the $\phi\to K^+K^-\to\gamma (f_0+\sigma)$ decay amplitude,
 otherwise  we got
$BR(\phi\to\gamma (f_0+\sigma)\to\gamma\pi\pi)\gtrsim4.5\cdot10^{-5}$,
see Table I. Forgetting the possible models, one can say that from
the $\pi\pi$ scattering results
 $BR(\phi\to\gamma (f_0+\sigma)\to\gamma\pi\pi)
\gtrsim10^{-5}$, that is enormous for the suppressed by the OZI rule decay.

We gratefully acknowledge discussions with M. Arpagaus, S.I. Eidelman and
J.A Thompson.

This research was supported in part by the Russian Foundation for Basic
Research, grants  94-02-05 188, 96-02-00 548 and  INTAS-94-3986.

\section{Appendix.}

The function $f_{\rho}(m)$, taking into account the interference of
the identical pions, see Fig.1, has the following integral presentation
\begin{eqnarray}
\label{funcrho}
&&f_{\rho}(m)=1-\frac{3}{8m^2p^2\omega}\int_{m_{\pi}}^{\tilde m}
Re\Biggl(\frac{D_{\rho}(m)}
{D_{\rho}(z)}\Biggr)\{y[(s-2m^2)(E_+-py)-2\omega(E_+-py)^2-\\ \nonumber
&&-2mm_{\pi}^2]+\frac{p}{2}(z^2+m^2-s)(1-y^2)\}zdz,
\end{eqnarray}
where 
\begin{eqnarray}
&&p=\sqrt{(s-(m-m_{\pi})^2)(s-(m+m_{\pi})^2)}/2m,\qquad 
\omega=(m^2-m_{\pi}^2)/2m,\\ \nonumber 
&&E_+=(s-m_{\pi}^2-m^2)/2m,\quad z^2=m_{\pi}^2+2\omega(E_+-py),\quad
\tilde m^2=2m_{\pi}^2+2m_{\phi}\omega-m^2.
\end{eqnarray}

The function $f_{\omega}(m)$ is resulted from (\ref{funcrho}) by replacement
$D_{\rho}(m)\to D_{\omega}(m)$.

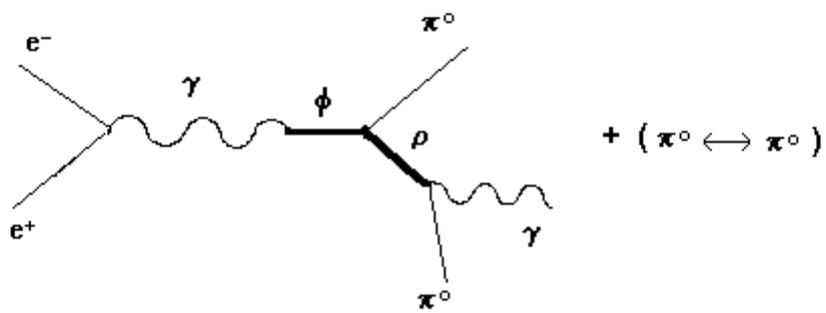
\begin{figure}
\caption{ The diagrams of the background to the $e^+e^-\to\gamma f_0\to
\gamma\pi^0\pi^0$ process in the vector dominance model . }
\end{figure}

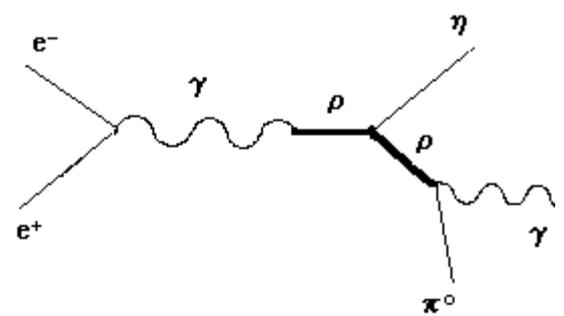
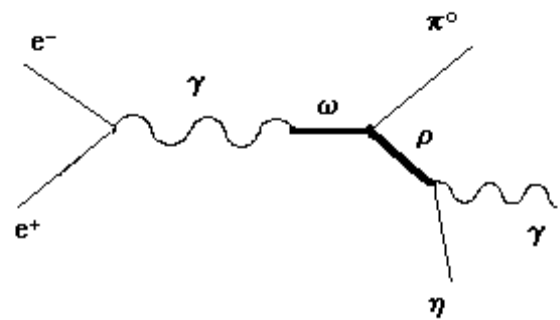
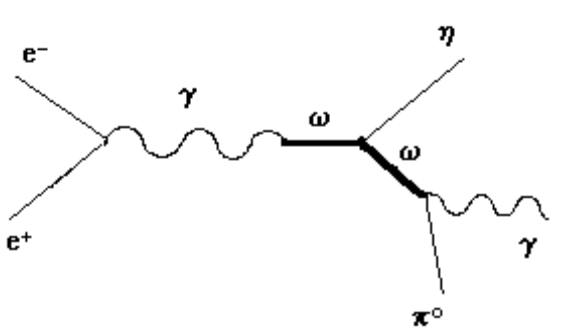
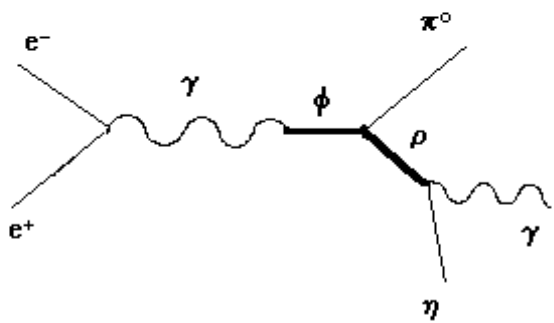
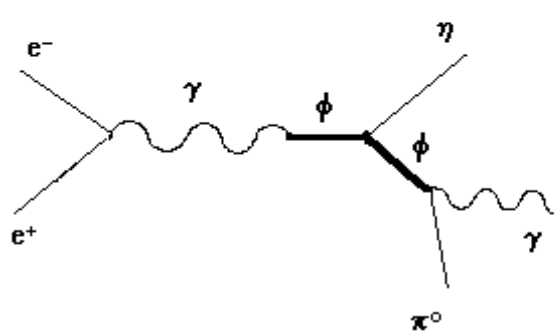
\begin{figure}
\caption{The diagrams of the background to the  $e^+e^-\to\gamma a_0\to
\gamma\pi^0\eta$ process in the vector dominance model. }
\end{figure}

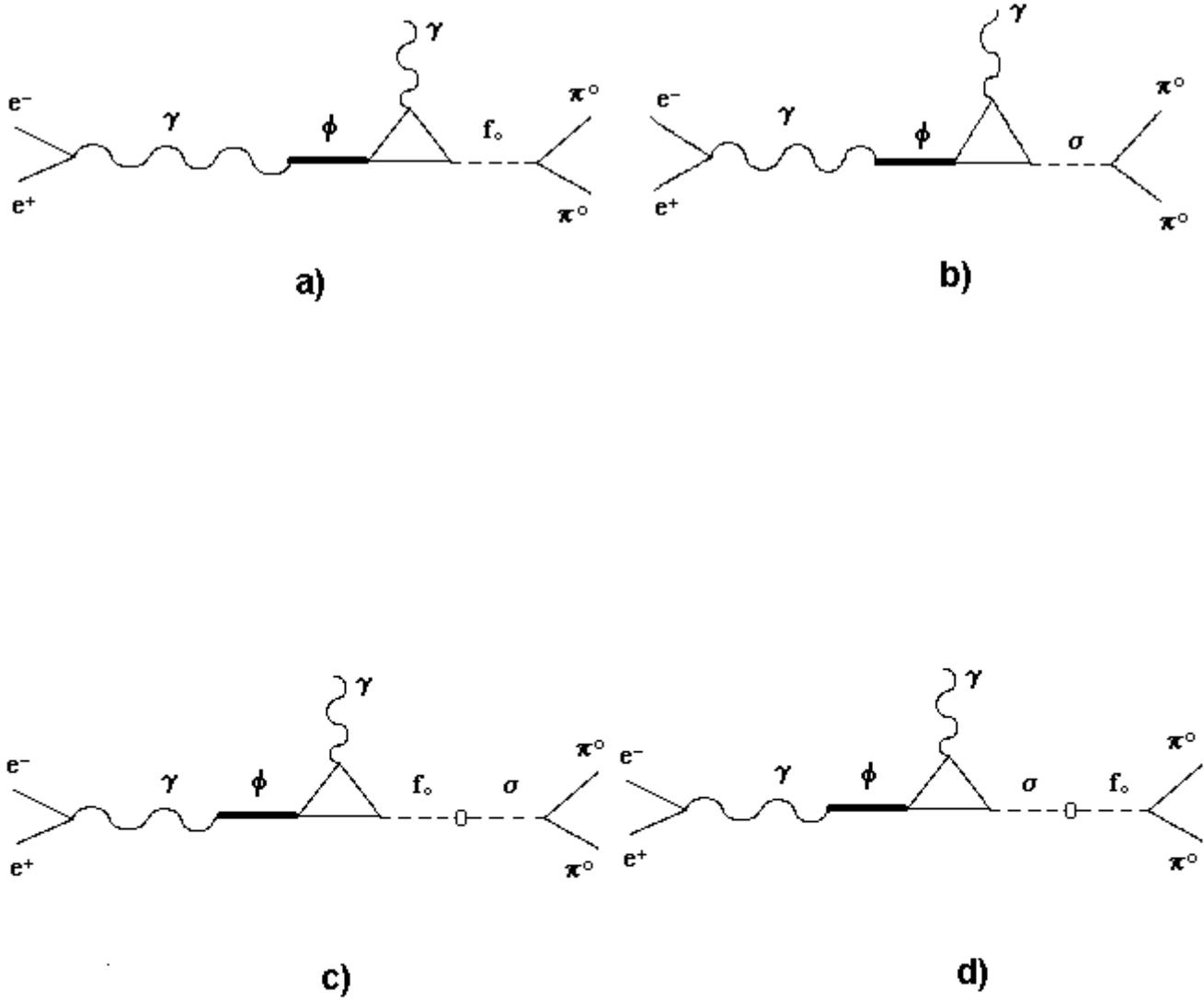
\begin{figure}
\caption{ The symbolic diagrams of the $e^+e^-\to\phi\to\gamma 
(f_0+\sigma)\to\gamma\pi\pi$ process with regard to the  $f_0$ and $\sigma$
mesons mixing. }
\end{figure}

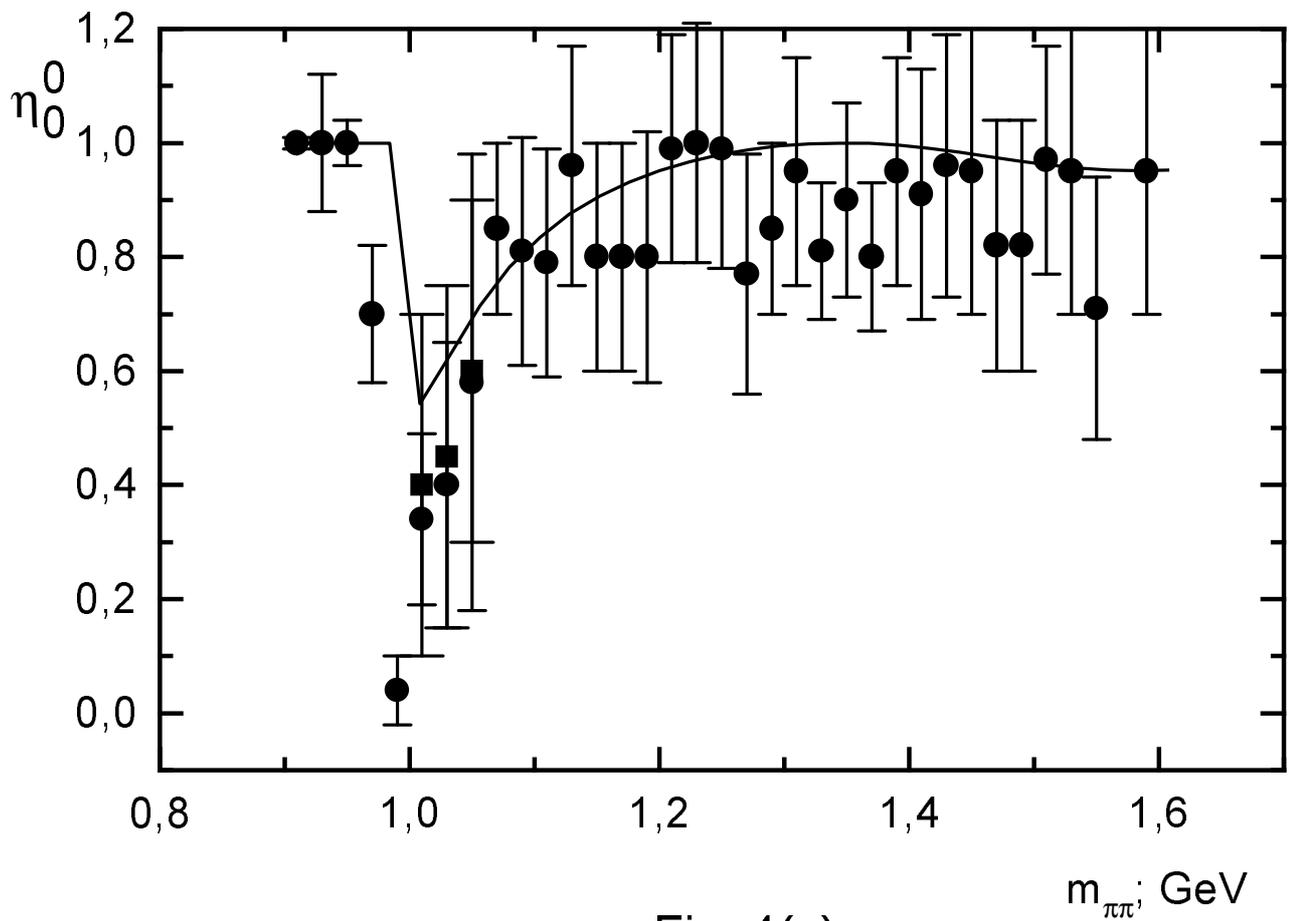
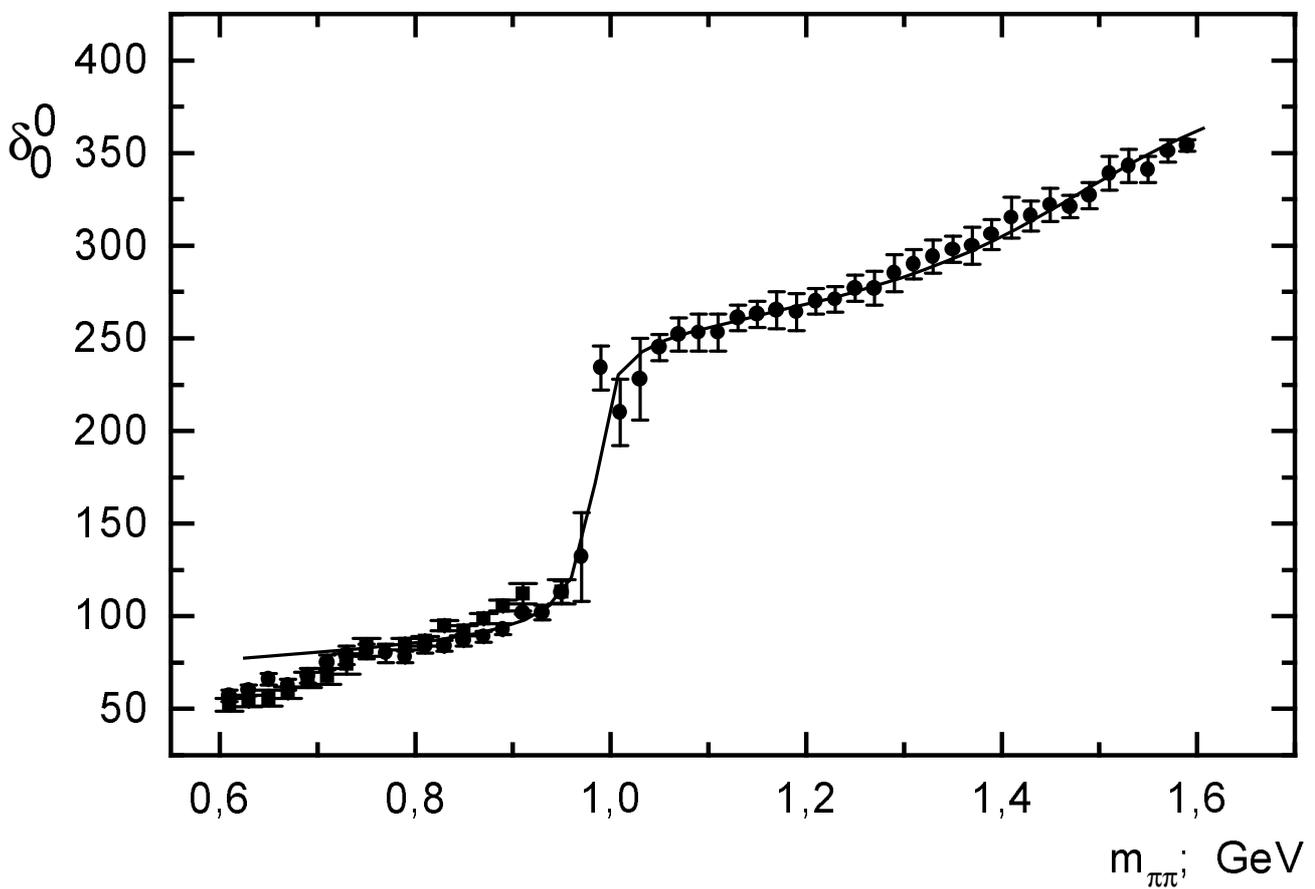
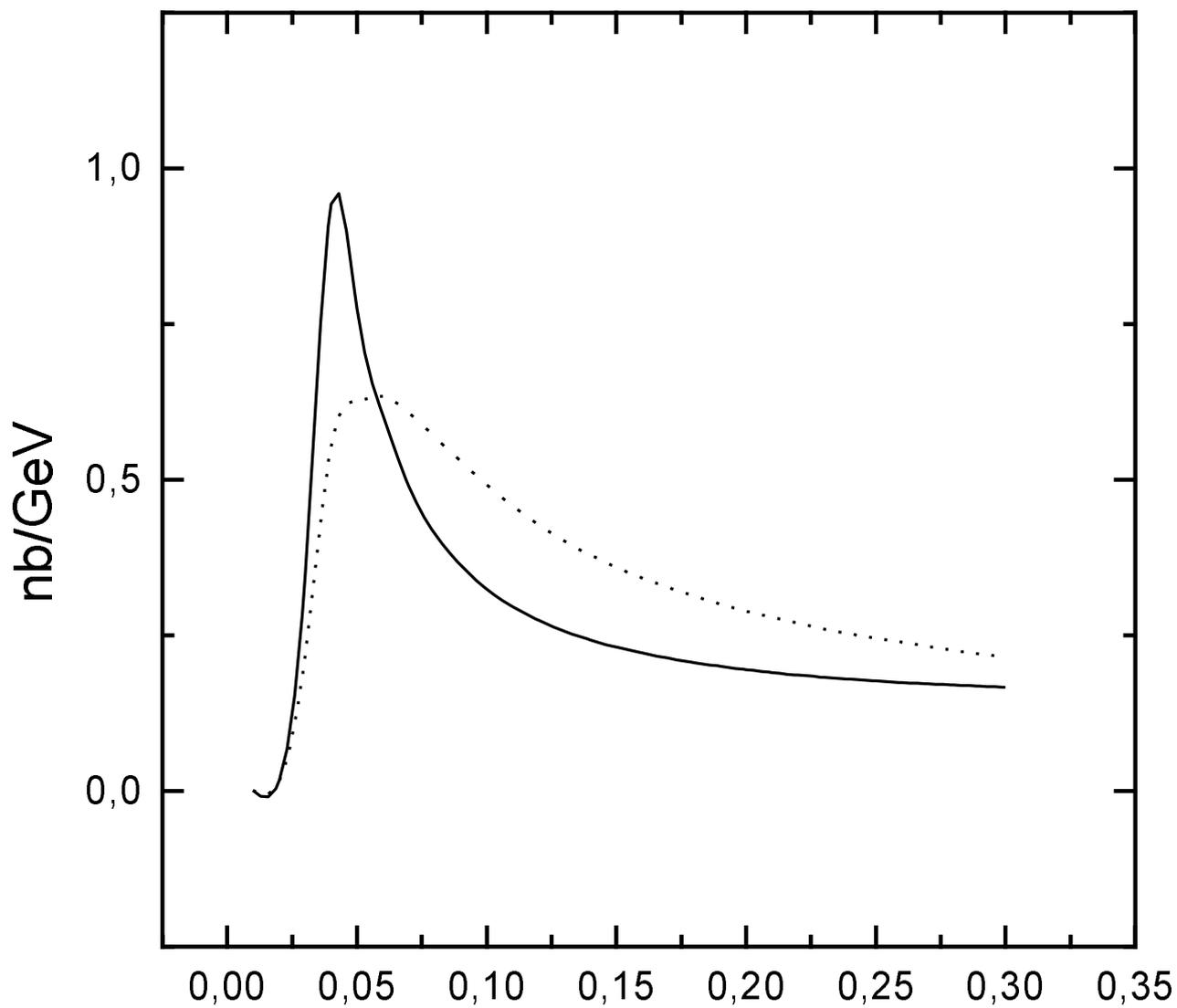
\begin{figure}
\caption{ The results of fitting in the $q^2\bar q^2$ model for the 
parameters:
$\theta=60^{\circ}$, $R=2.0$, $g^2_{f_0K^+K^-}/4\pi=0.72\ GeV^2$,
$g^2_{\sigma\pi\pi}/4\pi=1.76\ GeV^2$, $C_{f_0\sigma}=-0.17\ GeV^2$,
$m_{\sigma}=1.47\ GeV$, $m_{f_0}=0.98\ GeV$. The effective width of the
 $f_0$ meson $\Gamma_{eff}\simeq60\ MeV$. The $BR_{f_0+\sigma}(BR_{f_0})=
3\cdot BR(\phi\to\gamma(f_0+\sigma)\to\gamma\pi^0\pi^0)(
3\cdot BR(\phi\to\gamma f_0\to\gamma\pi^0\pi^0))=1.18(1.43)\cdot10^{-4}$
at $\omega<250\ MeV$ and  $BR_{f_0+\sigma}(BR_{f_0})=0.63(0.6)\cdot10^{-4}$
at $\omega<100\ MeV$.  (a) The inelasticity $\eta^{I=0}_{L=0}$. (b) 
The phase $\delta^{I=0}_{L=0}$. (c) The spectrum of the differential cross
section $d\sigma(e^+e^-\to\gamma(f_0+\sigma)\to\gamma\pi^0\pi^0)/d\omega$ 
with  mixing of the $f_0$ ang $\sigma$ mesons, see Eq.(\ref{signal}).  
The dashed line is the spectrum of the $f_0$ meson without mixing with
the $\sigma$ meson. The cut of the angle of the photon direction
$c=0.66$ reduces the cross section by 43\%. }
\end{figure}

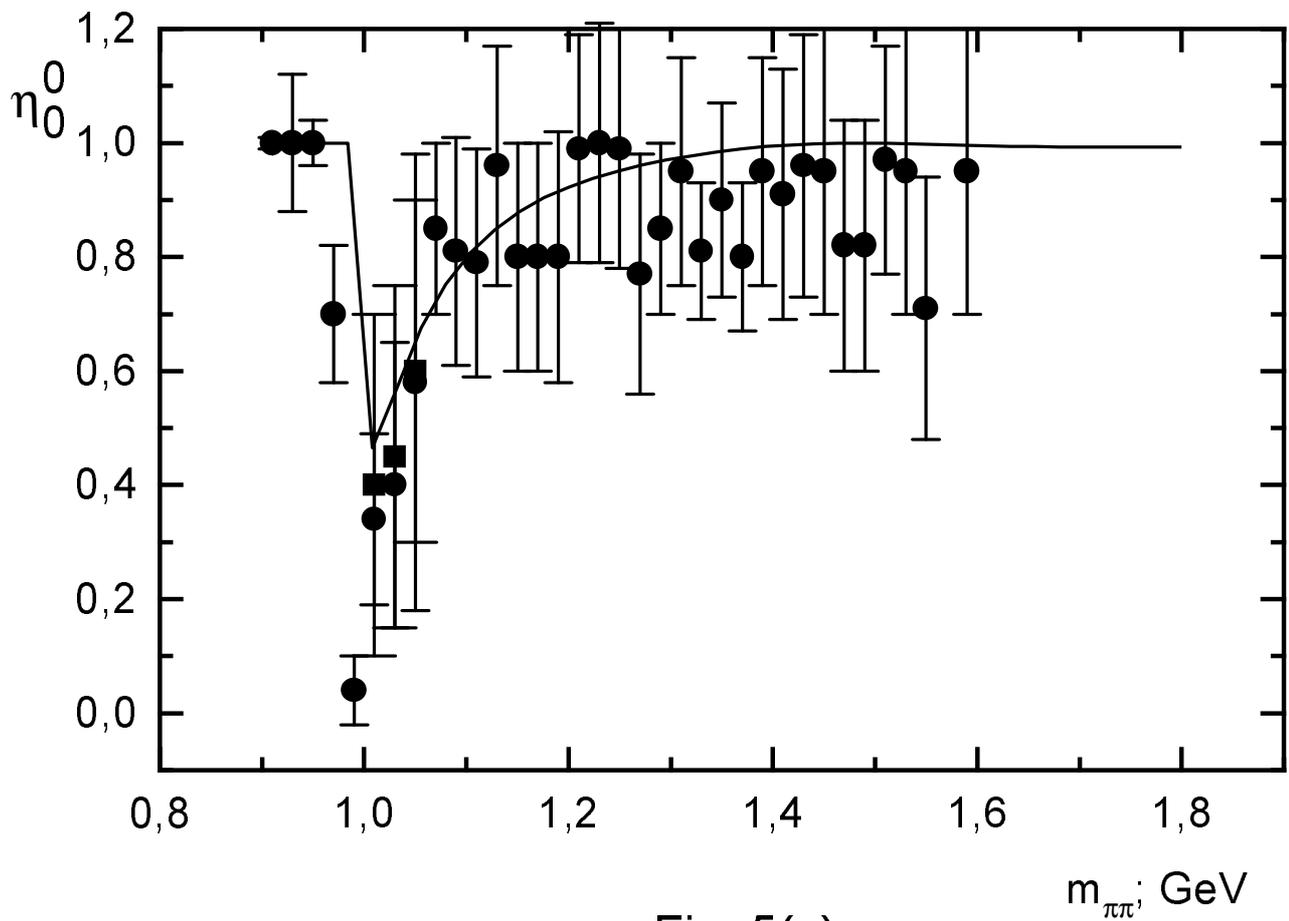
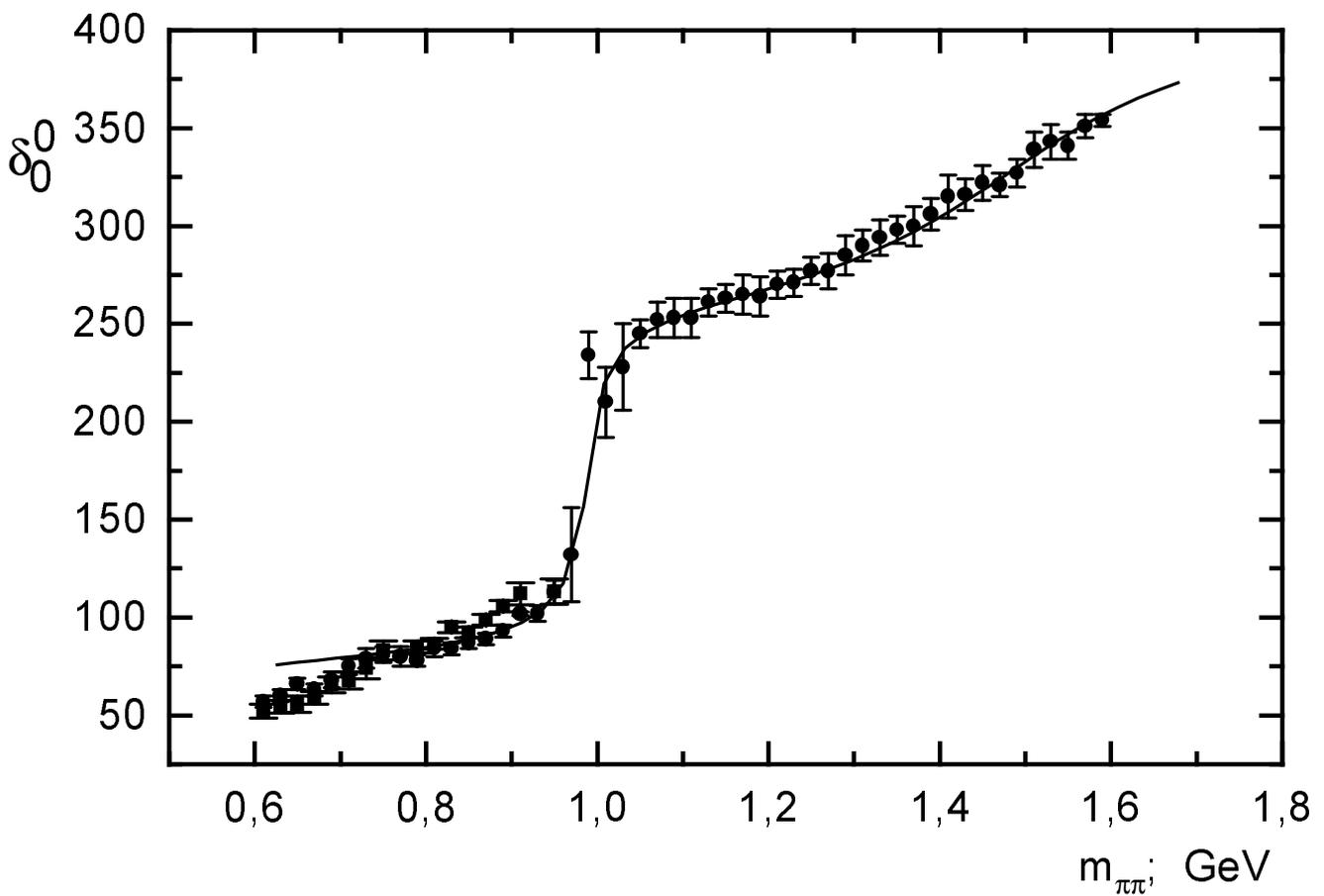
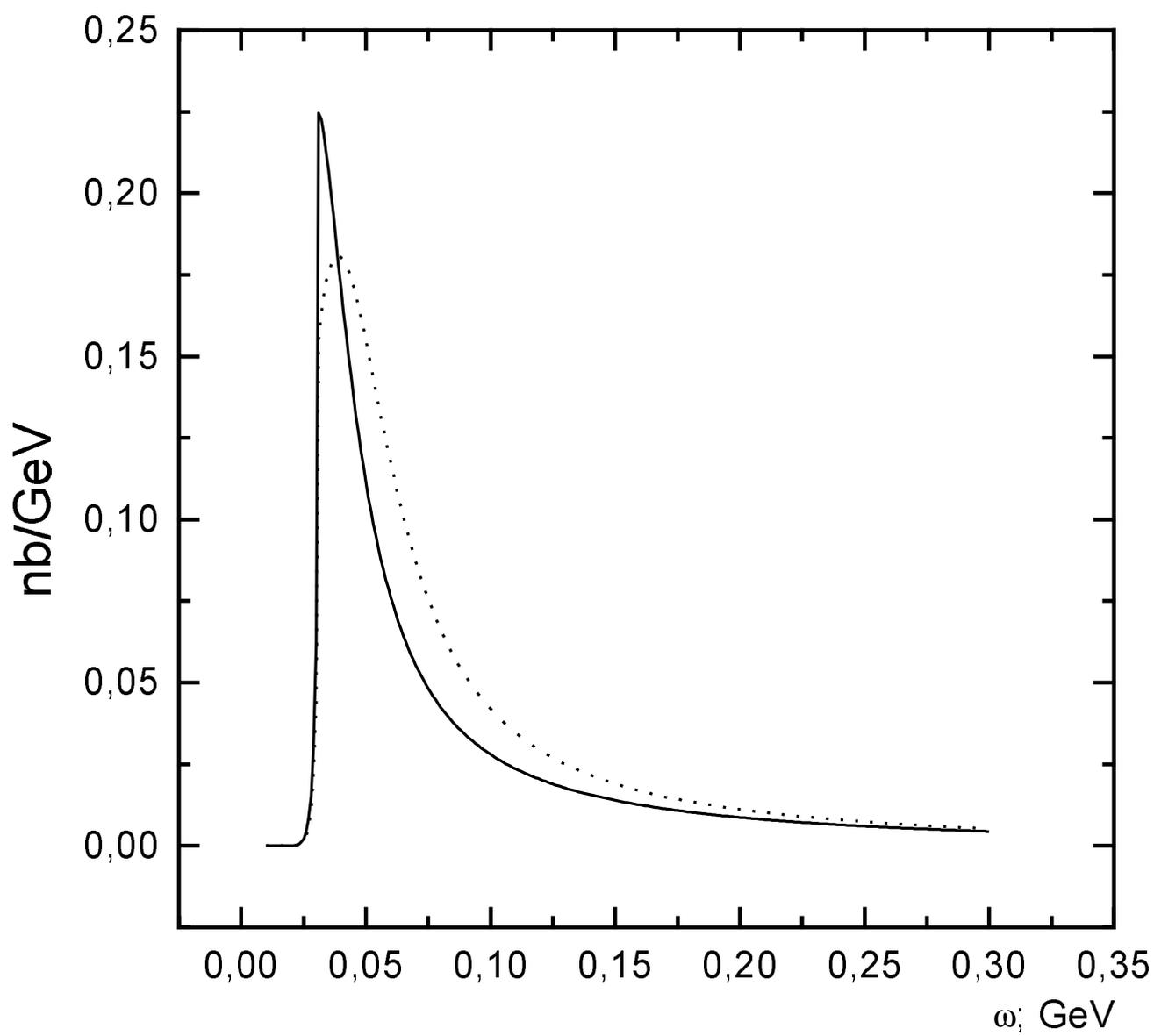
\begin{figure}
\caption{ The results of fitting in the model of the $K\bar K$ molecule
for the parameters:
$\theta=58^{\circ}$, $\Gamma_0=0.1\ GeV$, $g^2_{f_0K^+K^-}/4\pi=0.6\ GeV^2$,
$g^2_{\sigma\pi\pi}/4\pi=2.15\ GeV^2$, $m_{\sigma}=1.48\ GeV$,
 $m_{f_0}=0.98\ GeV$. The effective width of the $f_0$ meson
$\Gamma_{eff}\simeq30\ MeV$. The $BR_{f_0+\sigma}(BR_{f_0})=
3\cdot BR(\phi\to\gamma(f_0+\sigma)\to\gamma\pi^0\pi^0)(
3\cdot BR(\phi\to\gamma f_0\to\gamma\pi^0\pi^0))=1.21(1.53)\cdot10^{-5}$
at $\omega<250\ MeV$ and
$BR_{f_0+\sigma}(BR_{f_0})=0.91(1.13)\cdot10^{-5}$ at $\omega<100\ MeV$.
(a) The inelasticity $\eta^{I=0}_{L=0}$. (b) The phase $\delta^{I=0}_{L=0}$.
 (c) The spectrum of the differential cross section
 $d\sigma(e^+e^-\to\gamma(f_0+\sigma)\to\gamma\pi^0\pi^0)/d\omega$ 
with the moxing of the  $f_0$ and $\sigma$ mesons, see Eq.(\ref{signal}). 
The dashed line is the spectrum of the $f_0$ meson without mixing with the
$\sigma$ meson. The cut of the angle of the photon derection
$c=0.66$ reduces the cross section by 43\%. }
\end{figure}

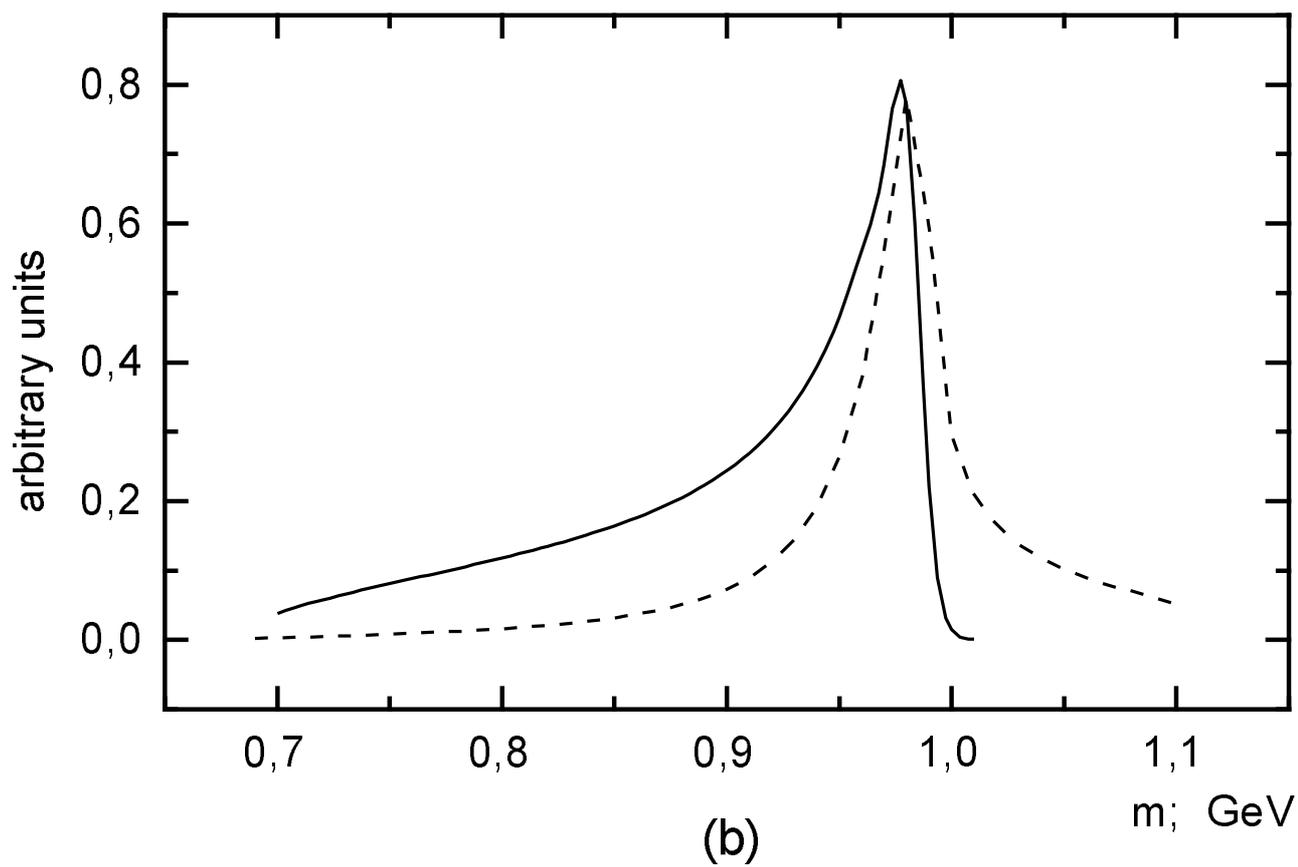
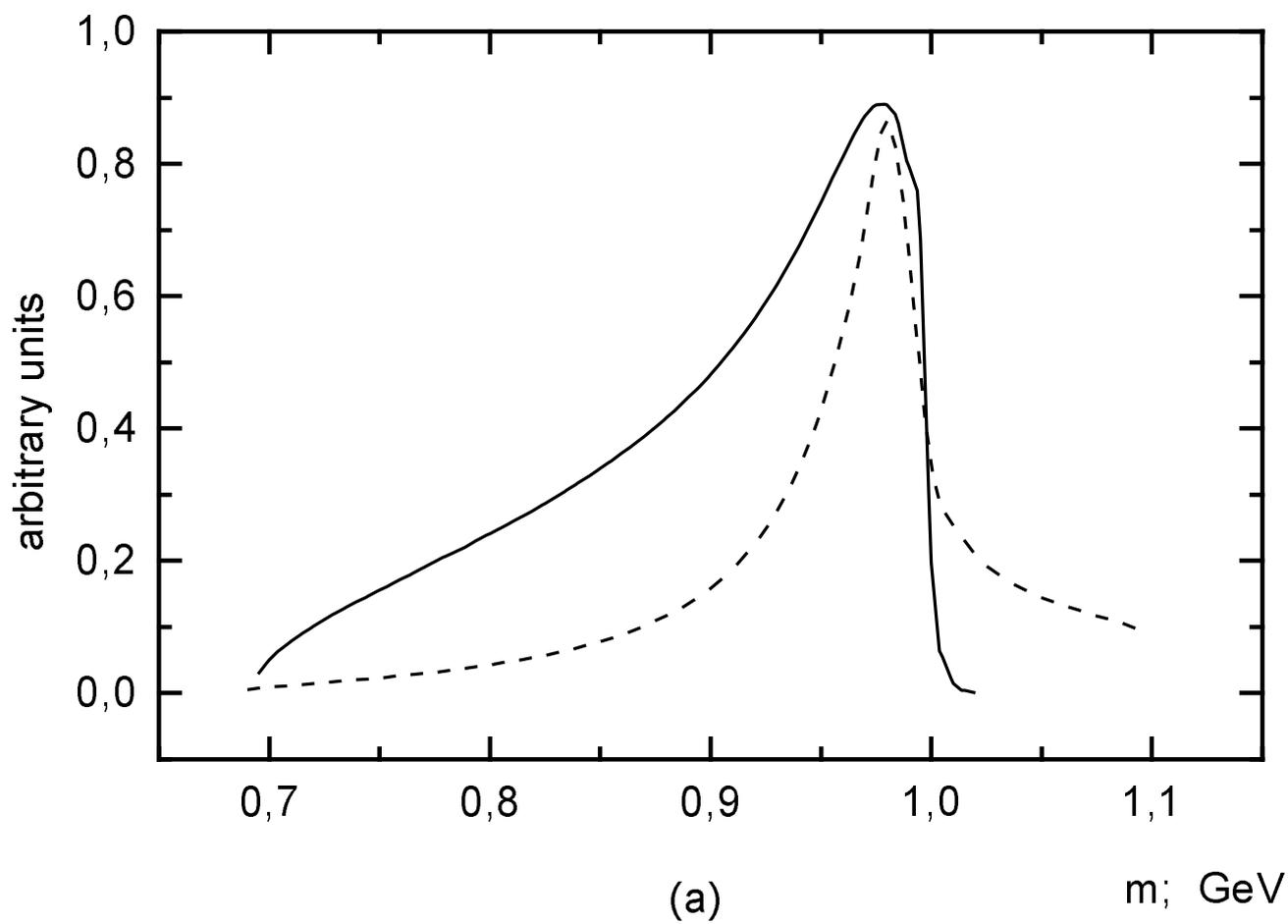
\begin{figure}
\caption{ The mass spectrum of the $\pi\eta$ system in the 
$e^+e^-\to\phi\to\gamma a_0\to\gamma\pi\eta$ reaction in the  $q^2\bar q^2$
model. The dashed line is the spectrum  $dN/dm\sim m^2\Gamma_{\pi\eta}
/|D_{a_0}(m)|^2$. (a) The constant  $g^2_{a_0K^+K^-}/4\pi=1.47\ GeV^2$, 
 $m_{a_0}=0.98\ GeV$.
The effective width of the  $a_0$ meson $\Gamma_{eff}\simeq60\ MeV$.
(b) The constant  $g^2_{a_0K^+K^-}/4\pi=0.72\ GeV^2$,  $m_{a_0}=0.985\ GeV$.
The effective width of the $a_0$ meson $\Gamma_{eff}\simeq 34\ MeV$.}
\end{figure}

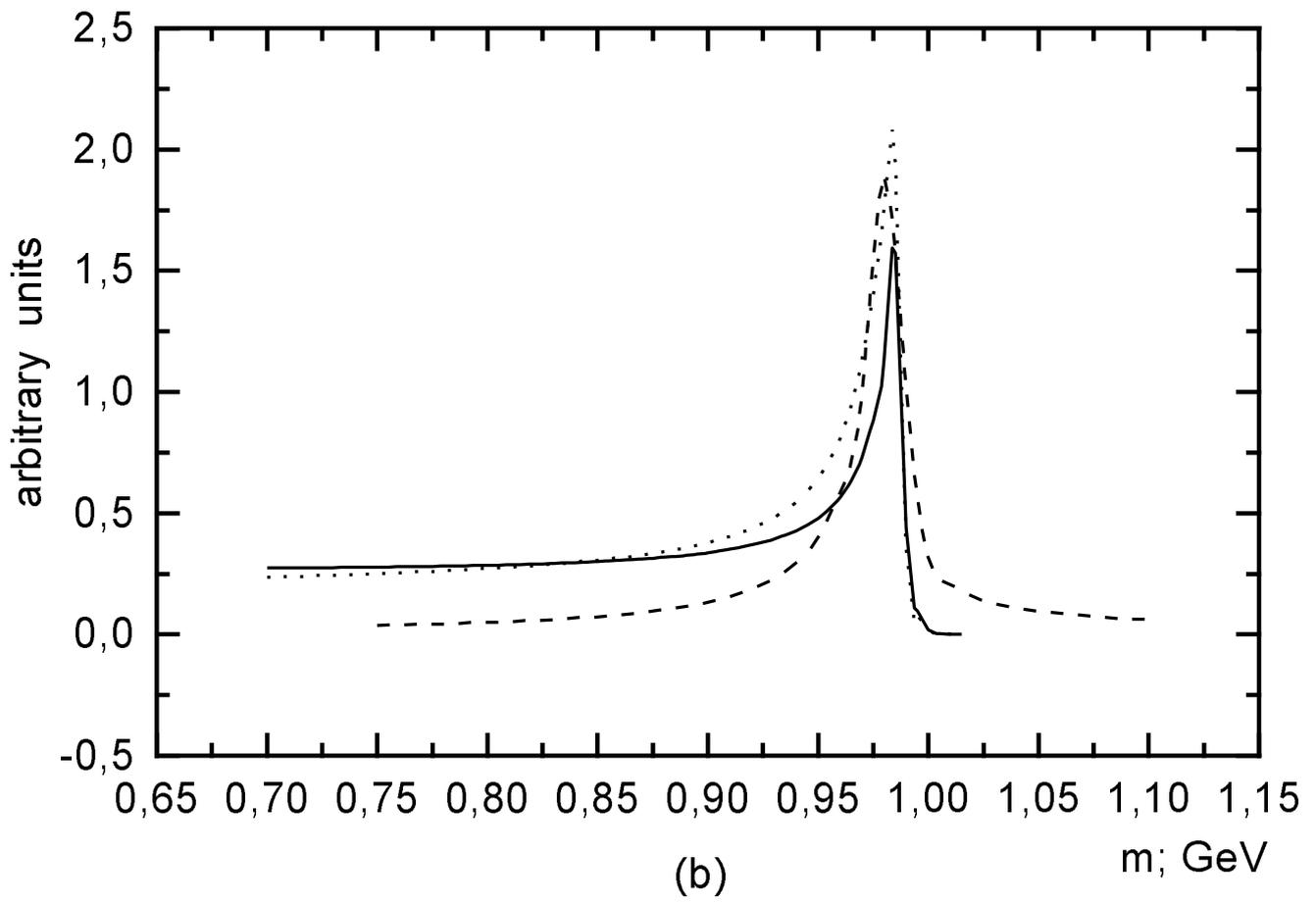
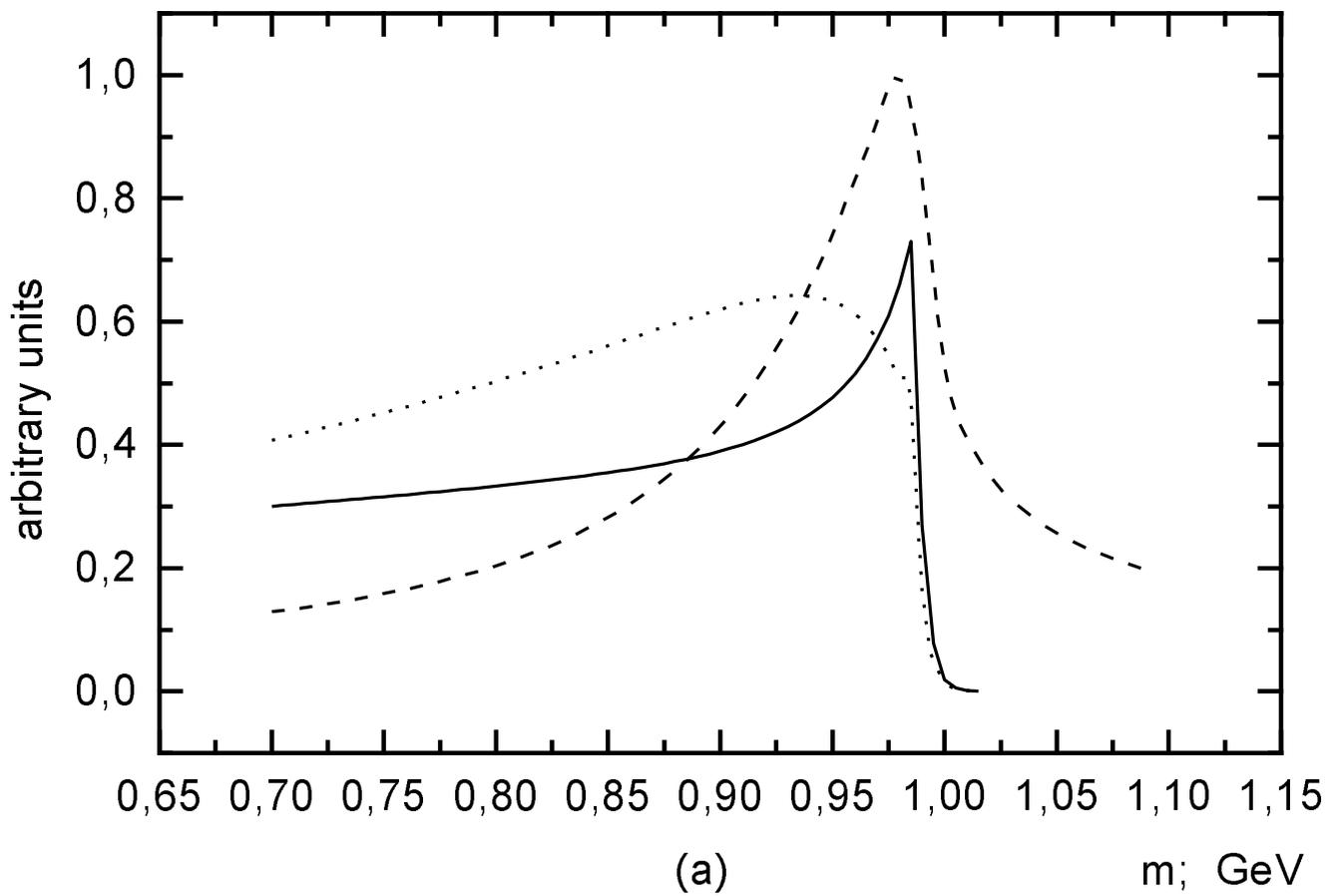
\begin{figure}
\caption{ The mass spectrum of the $\pi\pi$ system in the 
$e^+e^-\to\phi\to\gamma (f_0+\sigma)\to\gamma\pi\pi$ reaction in the
 $q^2\bar q^2$ model. The dotted line is the mass spectrum without mixing
with the $\sigma$ meson. The dashed line is the spectrum
 $dN/dm\sim m^2\Gamma_{\pi\pi}/|D_{f_0}(m)|^2$.
(a) $\theta=45^{\circ}$, $R=2$, $g^2_{f_0K^+K^-}/4\pi=1.47\ GeV^2$,
$g^2_{\sigma\pi\pi}/4\pi=1.76\ GeV^2$, $C_{f_0\sigma}=-0.31\ GeV^2$,
$m_{\sigma}=1.38\ GeV$, $m_{f_0}=0.985\ GeV$. The effective width of the
 $f_0$ meson $\Gamma_{eff}\simeq85\ MeV$.
(b)  $\theta=60^{\circ}$, $R=8$, $g^2_{f_0K^+K^-}/4\pi=2.25\ GeV^2$,
$g^2_{\sigma\pi\pi}/4\pi=1.76\ GeV^2$, $C_{f_0\sigma}=-0.31\ GeV^2$,
$m_{\sigma}=1.47\ GeV$, $m_{f_0}=0.98\ GeV$. The effective width of the
 $f_0$ meson  $\Gamma_{eff}\simeq25\  MeV$. }
\end{figure}

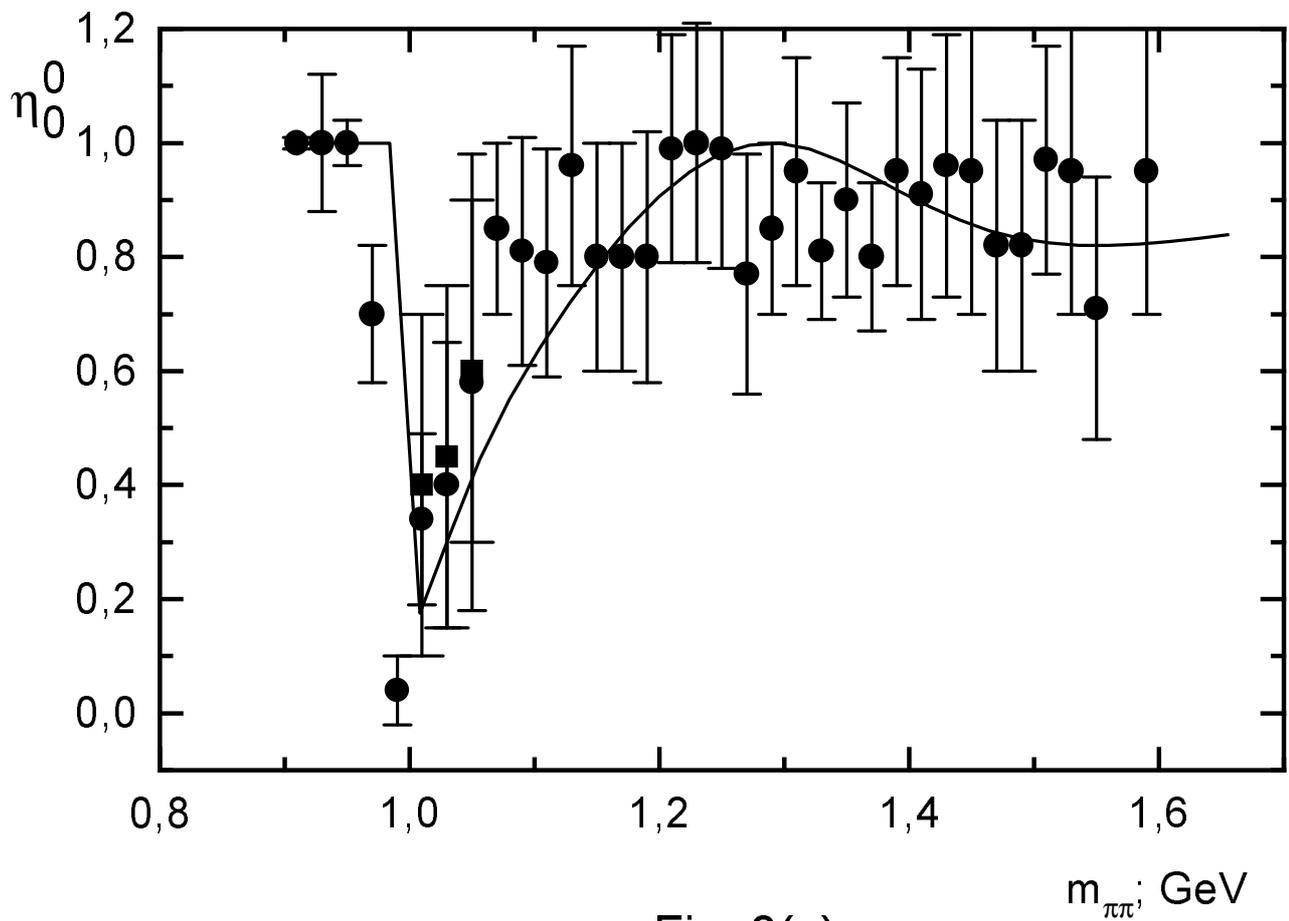
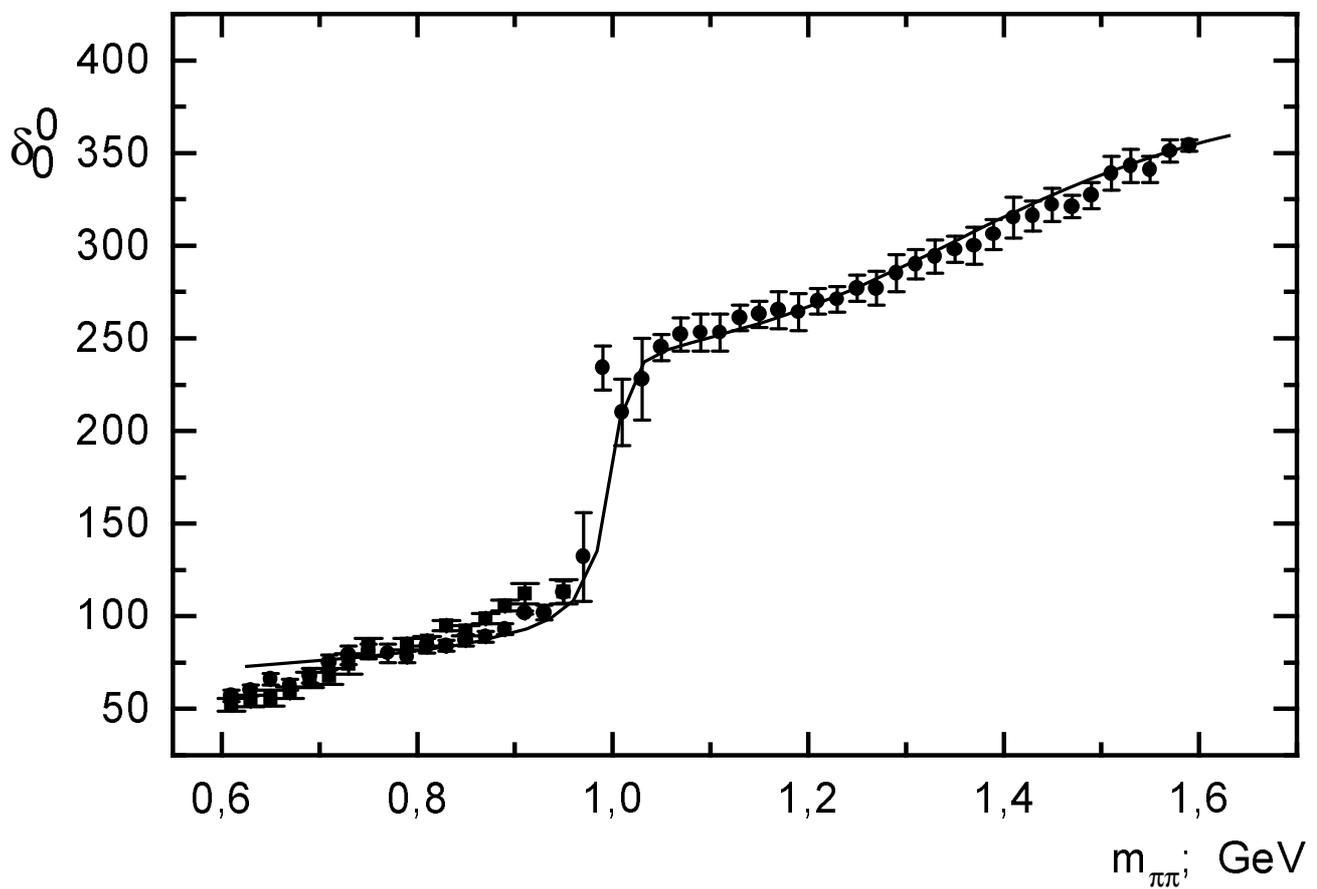
\begin{figure}
\caption{ The phase and inelasticity of the $\pi\pi$ scattering for the same
parametres as in Fig.7(a).}
\end{figure}

\begin{table}
\caption{ The results in the  $q^2\bar q^2$ model.
In the three last lines the parameters and results are listed for the
 $s\bar s$ model of the $f_0$ meson. All demensional quantities are
shown in units of GeV or $GeV^2$.
 The details are described in the Sec.IV .}

\vspace*{0.5cm}

\begin{tabular}{|c|c|c|c|c|c|c|c|c|c|}

$\theta$&R& $\frac{g^2_{f_0K^+K^-}}{4\pi}$&$\frac{g^2_{\sigma\pi\pi}}{4\pi}$&
$m_{\sigma}$&$-C_{f_0\sigma}$&$\Gamma_{eff}$&$BR_{f_0+\sigma}
(BR_{f_0})\cdot10^4$&$BR_{f_0+\sigma}(BR_{f_0})\cdot10^4;\omega<0.1$&
$m_{f_0}$ \\ \hline

$45^{\circ}$
&2.0&1.47&1.76&1.38&0.31&0.095&1.05(0.95)&0.71(0.76)&0.98\\ \hline
$45^{\circ}$
&2.0&1.47&1.76&1.38&0.31&0.100&1.21(1.14)&0.81(0.91)& 0.975\\ \hline
$45^{\circ}$
&2.0&1.47&1.76&1.38&0.31&0.110&2.36(3.0)&0.9(0.83)&0.970\\ \hline
$45^{\circ}$
&2.0&1.47&1.76&1.38&0.31&0.085&1.67(2.32)&0.59(0.68)&0.985\\ \hline
$50^{\circ}$
&2.0&1.47&1.76&1.4 &0.28&0.090&1.41(2.00)&0.49(0.59)& 0.990\\ \hline
$60^{\circ}$
&2.0&0.72&1.76&1.47&0.17&0.060&1.18(1.43)&0.63(0.6)&0.98\\ \hline
$60^{\circ}$
&2.0&0.72&1.76&1.47&0.17&0.055&1.01(1.27)&0.52(0.54)& 0.985\\ \hline
$60^{\circ}$
&2.0&0.72&1.76&1.47&0.17&0.060&0.81(1.09)&0.39(0.45)& 0.990\\ \hline
$60^{\circ}$
&2.0&0.72&1.76&1.47&0.17&0.070&1.34(1.56)&0.73(0.66)& 0.975\\ \hline
$55^{\circ}$
&4.0&1.47&2.1&1.47&0.31&0.042&1.6(2.05)&0.76(0.97)& 0.980\\ \hline
$50^{\circ}$
&4.0&4.25&1.76&1.47&0.31&0.061&2.9(3.8)&1.24(1.22)& 0.980\\ \hline
$60^{\circ}$
&8.0&2.25&1.76&1.47&0.31&0.025&2.74(3.36)&1.0(0.79)&0.98\\ \hline
$60^{\circ}$
&8.0&2.25&1.91&1.49&0.31&0.017&2.39(3.04)&0.75(0.74)& 0.985\\ \hline
$60^{\circ}$
&8.0&2.25&1.91&1.49&0.31&0.030&3.06(3.63)&1.22(0.83)& 0.975\\ \hline
$60^{\circ}$
&9.0&0.72&2.0&1.48&0.29&0.015&1.11(1.43)&0.68(0.6)&0.98\\ \hline
$60^{\circ}$
&9.0&0.72&2.0&1.48&0.29&0.010&0.97(1.27)&0.54(0.54)& 0.985\\ \hline
$60^{\circ}$
&9.0&0.72&2.0&1.48&0.29&0.005&0.82(1.1)&0.38(0.45)& 0.990\\ \hline
$60^{\circ}$
&15.0 &1.47&2.0&1.48&0.35&0.010&1.94(2.57)&1.02(0.74)&0.98\\ \hline
\hline
\hline
$55^{\circ}$
&4.0 &0.3&2.2&1.45&0.29&0.02&0.45(0.52)&0.29(0.33)&0.985\\ \hline
$50^{\circ}$
&2.0 &0.3&2.2&1.43&0.29&0.04&0.52(0.6)&0.27(0.38)&0.98\\ \hline
$50^{\circ}$
&1.0 &0.3&2.2&1.43&0.29&0.06&0.52(0.6)&0.24(0.31)&0.98
\end{tabular}
\end{table}

\begin{table}
\caption{The parameters and results in the model of the $K\bar K$ molecule.
The $BR_{f_0+\sigma}(BR_{f_0})=
3\cdot BR(\phi\to\gamma(f_0+\sigma)\to\gamma\pi^0\pi^0)(
3\cdot BR(\phi\to\gamma f_0\to\gamma\pi^0\pi^0))$ are listed for the region
 $\omega<250\ MeV$ ( $0.7<m<m_{\phi}$ ). All demensional quantities are
shown in  units of GeV or $GeV^2$.
 The details are described in the Sec.IV.}
\vspace*{1cm}
\begin{tabular}{|c|c|c|c|c|c|c|c|c|}

$\theta$&$\Gamma_0$& $\frac{g^2_{f_0K^+K^-}}{4\pi}$&$\frac{g^2_{\sigma\pi\pi}}{4\pi}$&
$m_{\sigma}$&$\Gamma_{eff}$&$BR_{f_0+\sigma}(BR_{f_0})\cdot10^5$
&$BR_{f_0+\sigma}(BR_{f_0})\cdot10^5;\omega<0.1$&
$m_{f_0}$ \\ \hline
$58^{\circ}$&0.10&0.6&2.15&1.48&0.03&1.21(1.53)&0.91(1.13)&0.98\\ \hline
$58^{\circ}$&0.10&0.6&2.15&1.48&0.02&1.00(1.34)&0.73(0.98)&0.985\\ \hline
$58^{\circ}$&0.10&0.6&2.15&1.48&0.015&0.78(1.11)&0.55(0.79)&0.99\\ \hline
$58^{\circ}$&0.10&0.6&2.15&1.48&0.035&1.41(1.7)&1.1(1.25)&0.975\\ \hline
$60^{\circ}$&0.05&0.6&2.1&1.5&0.01&1.82(2.66)&1.53(2.27)&0.985 

\end{tabular}
\end{table}

\begin{table}
\caption{The branching ratios of the $\phi\to\gamma a_0\to\pi\eta$ decay
in the  $q^2\bar q^2$ model.
In the two last lines the parameters and results are listed for the
 $s\bar s$ model of the $f_0$ meson. All demensional quantities are
shown in units of GeV or $GeV^2$.
 The details are described in the Sec.IV .}

\vspace*{1cm}
\begin{tabular}{|c|c|c|c|c|}

 $\frac{g^2_{a_0K^+K^-}}{4\pi}$ & $\Gamma_{eff}$ &$BR_{a_0}\cdot10^4;\
\omega<0.25$ & $BR_{a_0}\cdot10^4;\ \omega<0.1$ & $m_{a_0}$ \\ \hline

1.47&0.056&1.78&0.86&0.98\\ \hline
1.47&0.062&2.0&0.96&0.975\\ \hline
1.47&0.067&2.22&1.05&0.97\\ \hline
1.47&0.047&1.535&0.74&0.985\\ \hline
1.47&0.051&1.25&0.58&0.99\\ \hline
0.72&0.039&1.05&0.65&0.98\\ \hline
0.72&0.034&0.9&0.55&0.985\\ \hline
0.72&0.037&0.72&0.42&0.99\\ \hline
0.72&0.043&1.19&0.75&0.975\\ \hline
2.25&0.067&2.3&0.96&0.98\\ \hline
2.25&0.056&2.0&0.83&0.985\\ \hline
2.25&0.075&2.6&1.06&0.975\\ \hline
1.78&0.04&1.78&0.86&0.98\\ \hline
\hline
\hline
0.3&0.049&0.079&0.074&0.985\\ \hline
0.3&0.05&0.085&0.08&0.98
\end{tabular}
\end{table}

\begin{table}
\caption{ The branching ratios of the  $\phi\to\gamma a_0\to\pi\eta$ decay
in the model of the $K\bar K$ molecule. All demensional quantities are
shown in units of GeV or $GeV^2$.
 The details are described in the Sec.IV .}
 
\vspace*{1cm}
\begin{tabular}{|c|c|c|c|c|c|}
$\frac{g^2_{a_0K^+K^-}}{4\pi}$ & $\Gamma_0$ & $\Gamma_{eff}$ &
$BR_{a_0}\cdot10^5;\ \omega<0.25$ & $BR_{a_0}\cdot10^5;\ \omega<0.1$ &
 $m_{a_0}$ \\ \hline
0.6&0.05&0.027&0.85&0.73&0.975\\ \hline
0.6&0.05&0.023&0.74&0.64&0.98\\ \hline
0.6&0.05&0.020&0.60&0.51&0.985\\ \hline
0.6&0.05&0.023&0.43&0.35&0.99
\end{tabular}
\end{table}

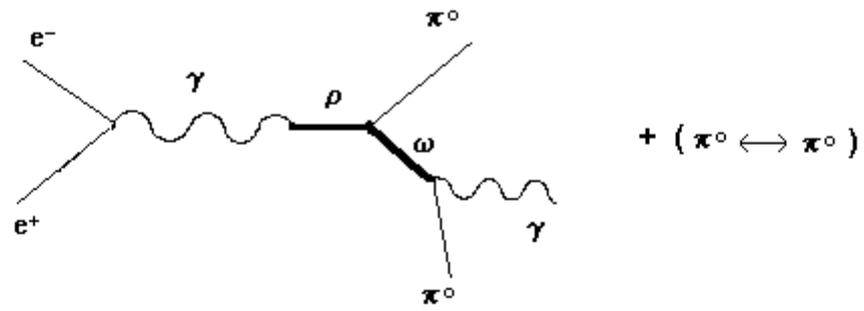
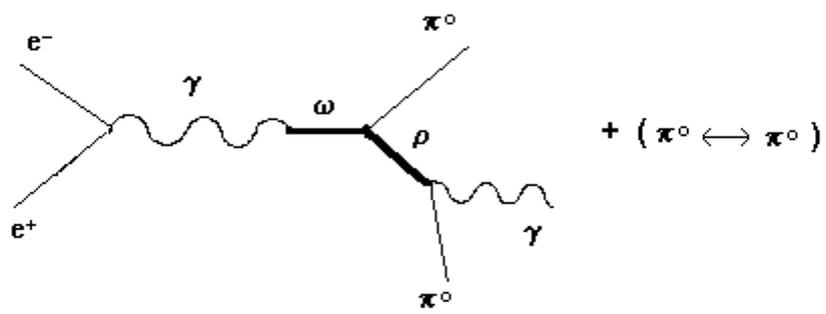
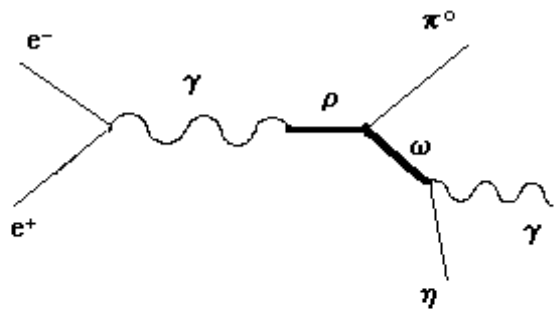
\end{document}